\def\includeappendix{yes}
\documentclass[
	aps,
	prx,
 amsmath,amssymb,
 superscriptaddress,
 reprint,
 showkeys,
]{revtex4-1}
\usepackage{afterpage}
\usepackage{siunitx}
\usepackage{graphicx}
\usepackage{dcolumn}
\usepackage{bm}
\usepackage[utf8]{inputenc}
\usepackage[T1]{fontenc}
\usepackage{mathptmx}
\usepackage{dsfont}
\usepackage{braket}
\usepackage{hyperref}
\usepackage{etoolbox}
\newcommand\xhat{\mathbf{\hat{x}}}
\newcommand\ii{\mathrm{i}}
\newcommand\zhat{\mathbf{\hat{z}}}
\newcommand\yhat{\mathbf{\hat{y}}}
\newcommand\rhat{\mathbf{\hat{r}}}

\newcommand\phir{\hat{\mathbf{\phi}}(\rhat)}
\newcommand\thetar{\hat{\mathbf{\theta}}(\rhat)}
\newcommand\hathelminuslambdar{{\boldsymbol\hat{\mathrm{\mathbf{e}}}_{-\lambda}}(\rhat)}
\newcommand\hathellambdar{{\boldsymbol\hat{\mathrm{\mathbf{e}}}_{\lambda}}(\rhat)}
\newcommand\dd{\mathbf{p}}
\newcommand\rr{\mathbf{r}}
\newcommand\alphahat{\mathbf{\hat{\alpha}}}
\newcommand\JJ{\mathbf{J}}
\newcommand\PP{\mathbf{P}}
\newcommand\FF{\mathbf{F}}
\newcommand\Mx{\mathrm{M}_{\mathrm{x}}}
\newcommand\Mz{\mathrm{M}_{\mathrm{z}}}
\newcommand\restr[2]{{
  \left.\kern-\nulldelimiterspace 
  #1 
  \vphantom{\big|} 
  \right|_{#2} 
  }}
\newcommand\Eq[1]{Eq.~(\ref{#1})}

\newcommand\duetoeq[1]{\stackrel{\text{Eq.\ (\ref{#1})}}{=}}
\newcommand\equaldueto[1]{\stackrel{#1}{=}}
\newcommand\sigmapm{\ket{\sigma_{\pm \hat{\mathrm{x}}}}}
\newcommand\mplambda{\ket{\mathrm{k} \ j\ m_z\ \lambda}}
\newcommand\mplambdamy{\ket{\mathrm{k} \ j\ m_y\ \lambda}}
\newcommand\mplambdaplus{\ket{\mathrm{k} \ j\ m_z\ \lambda=+1}}
\newcommand\mplambdaminus{\ket{\mathrm{k} \ j\ m_z\ \lambda=-1}}
\newcommand\mptau{\ket{\mathrm{k} \ j\ m_z\ \tau}}
\newcommand\mptauplus{\ket{\mathrm{k} \ j\ m_z\ \tau=+1}}
\newcommand\mptauminus{\ket{\mathrm{k} \ j\ m_z\ \tau=-1}}
\newcommand\sigmapmbra{\bra{\sigma_{\pm \hat{\mathrm{x}}}}}
\newcommand\sigmaplusbra{\bra{\sigma_{+ \hat{\mathrm{x}}}}}
\newcommand\sigmaminusbra{\bra{\sigma_{- \hat{\mathrm{x}}}}}
\newcommand\psiket{|\Psi\rangle}
 \ifdefstring{\includeappendix}{no}{\usepackage{xr}\externaldocument{main_appendices_transverse_couplings}}{}

\begin{document}

\title{Directional coupling of emitters into waveguides: A symmetry perspective}

\author{Aristeidis G. Lamprianidis}
 \email{aristeidis.lamprianidis@kit.edu}
 \affiliation{Institute of Theoretical Solid State Physics, Karlsruhe Institute of Technology, 76128 Karlsruhe, Germany}
 \author{Xavier Zambrana-Puyalto}
 \affiliation{Istituto Italiano di Tecnologia, Via Morego 30, 16163 Genova, Italy}
\author{Carsten Rockstuhl}
\affiliation{Institute of Theoretical Solid State Physics, Karlsruhe Institute of Technology, 76128 Karlsruhe, Germany}
\affiliation{Institute of Nanotechnology, Karlsruhe Institute of Technology, 76021 Karlsruhe, Germany}
\author{Ivan Fernandez-Corbaton}
 \email{ivan.fernandez-corbaton@kit.edu}
\affiliation{Institute of Nanotechnology, Karlsruhe Institute of Technology, 76021 Karlsruhe, Germany}

\date{\today}
\begin{abstract}
	Recent experiments demonstrated strongly directional coupling of light into waveguide modes. We identify here the mechanisms behind this effect. We consider emitters near a waveguide, either centered on the median plane of the waveguide, or displaced from such plane. We show that, independently of the displacement, the directionality is mostly due to a mirror symmetry breaking caused by the axial character of the angular momentum of the emitted light. The sign of the angular momentum along an axis transverse to the waveguide determines the preferential coupling direction. The degree of directionality grows exponentially as the magnitude of such transverse angular momentum increases linearly. We trace this exponential dependence back to a property of the evanescent angular spectrum of the emissions. A binary and less pronounced directional coupling effect due to the chiral character of the handedness of the emission is possible when the displacement of the emitter breaks another of the mirror symmetries of the waveguide. We find a selection rule that allows or prevents the coupling of centered electric(magnetic) multipolar emissions onto the waveguide modes. We also show that the selection of a different angular momentum axis made in some experiments causes significant differences in the way in which directionality depends on angular momentum. We then use these differences to propose an experiment featuring a transverse magnetic bias that allows to aggregate the directional emissions from quantum dots on top of waveguides. Our symmetry-based results apply to any emitted multipolar order, clarify the spin-momentum locking concept, and generalize it to an exponentially-strong locking between the transverse angular momentum and the preferential coupling direction. 
\end{abstract} 

\maketitle
\section{Introduction and summary}
Several recent experiments have demonstrated directional coupling of light into waveguide modes. For example, pronounced directionality has been shown in the collection of atomic emissions by optical fibers \cite{Mitsch2014b} and quantum dot emissions by waveguides \cite{Sollner2015,Coles2016,Scarpelli2019}. Similarly, experiments have shown pronounced directional coupling of focused light beams into waveguides, either directly \cite{Fang2019} or mediated by a scatterer \cite{Petersen2014,Rodriguez2014}. The directionality effect has the potential to route light and classify emissions according to the electromagnetic properties that determine the preferential coupling direction. Different theoretical approaches have been developed to understand the effect \cite{Le2014,Aiello2015,Bliokh2015,LeFeber2015,Bliokh2015b,transversecouplingevanescentwaves,EspinosaSoria2016,VanMechelen2016,Lodahl2017,Picardi2017,Wei2019,Savelev2019,Vazquez2019}. In particular, the concepts of transverse spin and spin-momentum locking in evanescent waves have been put forward as the origin of the directionality. Yet, a general and precise understanding is still lacking. For example, the dipolar approximation is routinely made to characterize the emitter. This precludes the study and prediction of possible directional coupling effects for the light emitted from higher-order multipolar transitions of atoms, molecules, and quantum dots \cite{Tojo2004,Andersen2011,Cheng2012,Karaveli2010,Kasperczyk2015,Vaskin2019}. Additionally, an ambiguity is introduced by the use of the photonic spin. In the context of the common separation of the optical angular momentum into orbital and spin parts\footnote{See e.g. the recent special issue and references therein \cite{Barnett2017}}, the spin of the photon is often simultaneously connected to both angular momentum and circular-polarization handedness (e.g. \cite{Oneil2002,Padgett2011,Cameron2012,Bliokh2015b}). This raises the question of which property dictates the directional coupling, since each of the two options implies fundamentally different characteristics and applications of the directional coupling effect. Finally, a static magnetic field or the incident excitation can be used in experiments to choose the axis of well-defined angular momentum for the emissions. In some experiments the choice has coincided with what we define as the transverse axis \cite{Mitsch2014b,Petersen2014,Rodriguez2014}. In other experiments, a different axis has been chosen \cite{Sollner2015,Coles2016,LeFeber2015,Scarpelli2019}. The theoretical description of both options is not yet unified in a single framework, and their potential differences regarding directionality have not been elucidated.

In this work, we will use a symmetry-based approach where the angular momentum is not separated into orbital and spin parts \cite{FerCor2012b,transversecouplingevanescentwaves}, and that successfully predicts some different effects that angular momentum and helicity can have in light-matter interactions \cite{Zambrana2016,Zambrana2018,Abdelrahman2019}. Using symmetry analysis and numerical simulations, we elucidate the mechanisms behind the directional coupling effect. Our approach and results are valid for emissions of a general multipolar order. In particular, we study the separate role of two different properties of the emission: Helicity(polarization handedness) and angular momentum. We consider emitters near a waveguide and either centered on the median plane of the waveguide, or displaced from such plane. We show that, independently of the displacement, the directionality is mostly determined by the projection of the angular momentum on the axis transverse to the plane defined by the position of the emitter and the waveguide axis. The directionality occurs because a particular mirror symmetry is broken due to the fact that the angular momentum is an axial vector. The sign of the transverse angular momentum vector determines the preferred coupling direction, while its magnitude determines both the degree of symmetry breaking and the degree of directionality, which grows with such magnitude. We show that such growth is {\em exponential}, and that this is due to an intrinsic characteristic of the evanescent components of the emissions, whose power flux in the relevant directions depends exponentially on the transverse angular momentum. The exponential dependence occurs for emissions of pure handedness as well as for their linear combinations, in particular electric and magnetic multipolar emissions, while coupling of electric(magnetic) emitters centered on the median plane of the waveguide is allowed or prevented by a selection rule. For centered emitters, the helicity of the emission has no effect in the directionality. Displacing the emitter allows a helicity dependent contribution to the directionality when the displacement breaks an additional mirror symmetry. This contribution, which is of significantly smaller magnitude than the one due to transverse angular momentum, biases the directionality in a position dependent way: For opposite displacements the same helicity produces an opposite bias. 

The transverse angular momentum axis has been experimentally selected in some cases e.g. \cite{Mitsch2014b,Petersen2014,Rodriguez2014}, while a different axis has been chosen in other cases e.g. \cite{Sollner2015,Coles2016,LeFeber2015,Scarpelli2019}. We elucidate the differences in directionality due to the two different choices. There are two salient differences. In the former case, centered emitters can show directionality and the sign of the directionality is locked to the sign of the angular momentum of the emission independently of the lateral displacement of the emitter. In the latter case, and in full agreement with the experiments, we provide explicit symmetry proofs that show that directionality is forbidden for emitters centered in mirror symmetric waveguides, and that the sign of the directionality depends on the sign of the lateral displacement. Then, we use these differences to propose an experiment that achieves the aggregation of the directional emissions of quantum dots on top of waveguides by using a magnetic bias in the transverse direction and illuminating all the quantum dots simultaneously.

According to our results, the dominant directionality effect due to transverse angular momentum could be exploited for routing light depending on its angular momentum, or for detecting high-order multipolar transitions of discrete emitters. Yet, it is not suited for applications that require handedness sensitivity, like discriminating between the two enantiomers of chiral molecules. In the rest of the article, we present the simulation results for the coupling directionality of an emitter near a rectangular waveguide and explain them using the broken and unbroken symmetries of the joint emission-waveguide system.
\begin{figure}
\includegraphics[width=8.5cm]{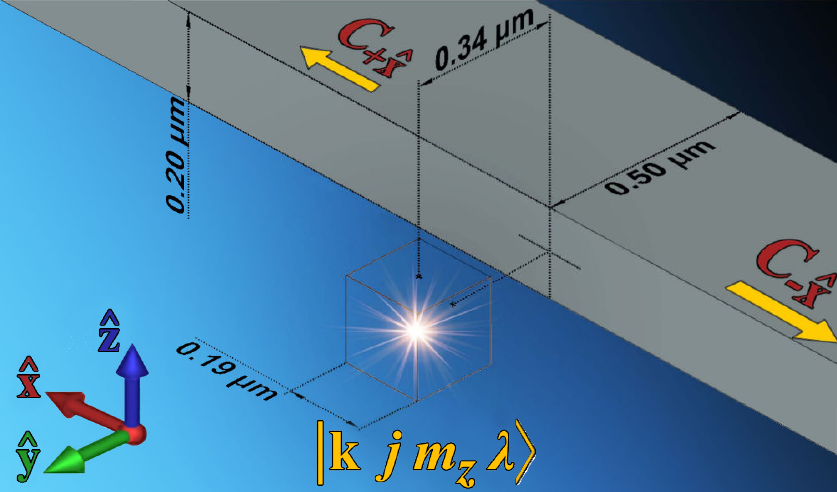}
	\caption{\label{fig:fig1} Sketch of the geometry of the system representing the coupling of the multipolar emission $\ket{\mathrm{k}\hspace{2pt}j\hspace{2pt}m_z\hspace{2pt}\lambda}$ to a nearby silicon waveguide. The emitter is located in vacuum at the center of the coordinate system. The waveguide is placed symmetrically with respect to the xOy plane with its optical axis parallel to the $x$-axis. The radiated power that couples to the first guided mode of the waveguide towards either the $+\hat{\mathrm{x}}$ or the $-\hat{\mathrm{x}}$ directions is collected by waveguide ports.}
\end{figure}

\section{Numerical simulations}
\begin{figure*}[ht]
\includegraphics[width=17cm]{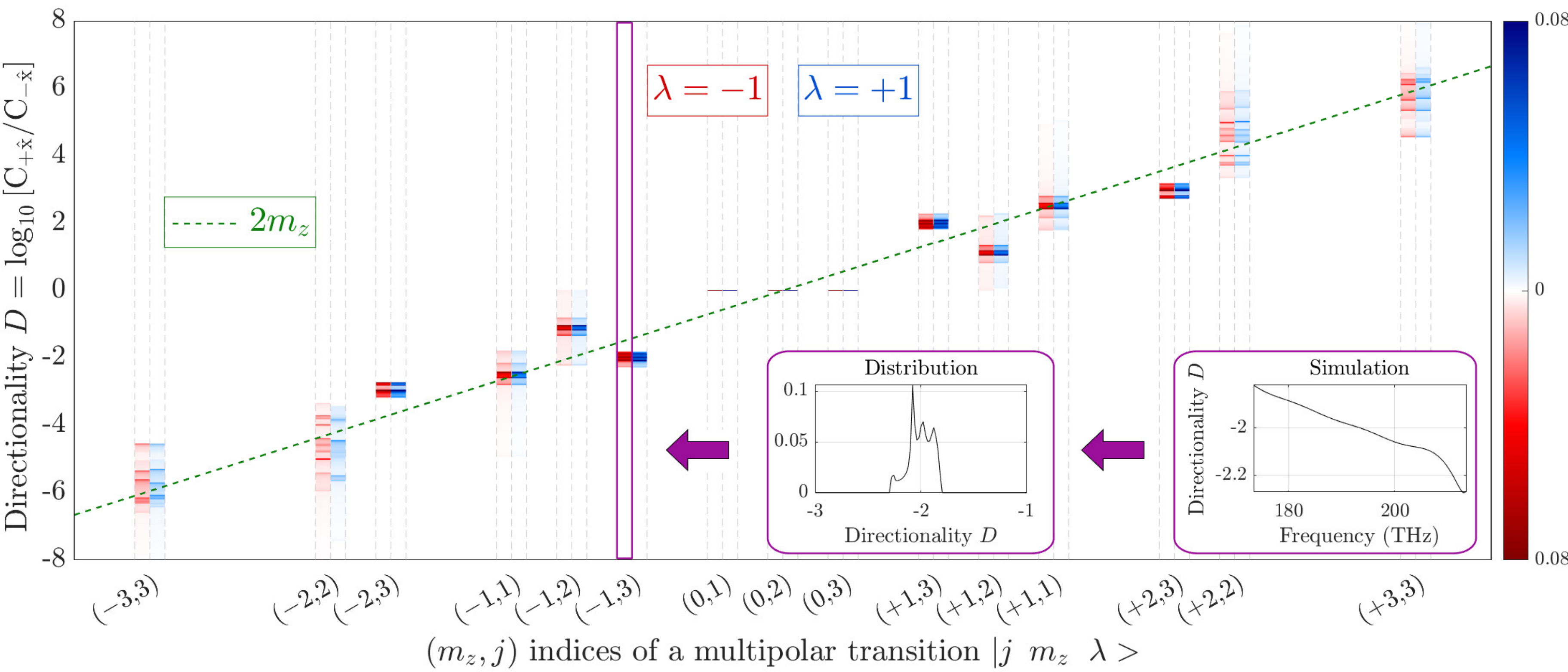}
	 \caption{\label{fig:fig2} For each $(j,m_z,\lambda)$, the graph shows the frequency distribution (see inset) of the logarithmic directionality of the coupling of the emitter into the waveguide. Blue(red) corresponds to multipolar emissions with positive(negative) helicity. The green dashed line corresponds to $2m_z$. Positive(negative) values of $D$ indicate preferential coupling towards the $+\hat{\mathrm{x}}(-\hat{\mathrm{x}})$ direction, and $|D|$ measures the degree of directionality in orders of magnitude. The graph shows how $D$ is mostly determined by the eigenvalue of the transverse component of the angular momentum, $m_z$. The independence of $D$ on the helicity $\lambda$ is clearly observed. }
\end{figure*}

Figure~\ref{fig:fig1} shows the considered geometry. An emitter is placed at the origin of the coordinate system close to a nearby rectangular silicon waveguide. The waveguide is invariant under reflections across the xOy and yOz planes, and is parallel to the $\mathrm{x}$-axis. The distance between the emitter and the axis of the waveguide is 590~nm. The width of the waveguide is 500~nm and its height is 200~nm. We perform numerical simulations over a frequency window of 40~THz centered at $f_0\approx193.4$~THz. The central frequency corresponds to a vacuum wavelength of $1550$~nm, and the frequency span to a wavelength range between $1404$~nm and $1729$~nm. For practical purposes, the waveguide can be considered single-mode across the entire frequency band\footnote{At the central frequency, the first waveguide mode has an effective index of about 2.26, whereas the effective index of its second mode is only about 1.07. This second very weakly guided mode will be practically irrelevant since any surface defect will efficiently scatter the guided power out of the waveguide.}. The simulations are performed in the time domain using CST MWS. Appendix~\ref{app:cst} contains detailed explanations.

\begin{figure}[h!]
\includegraphics[width=8.45cm]{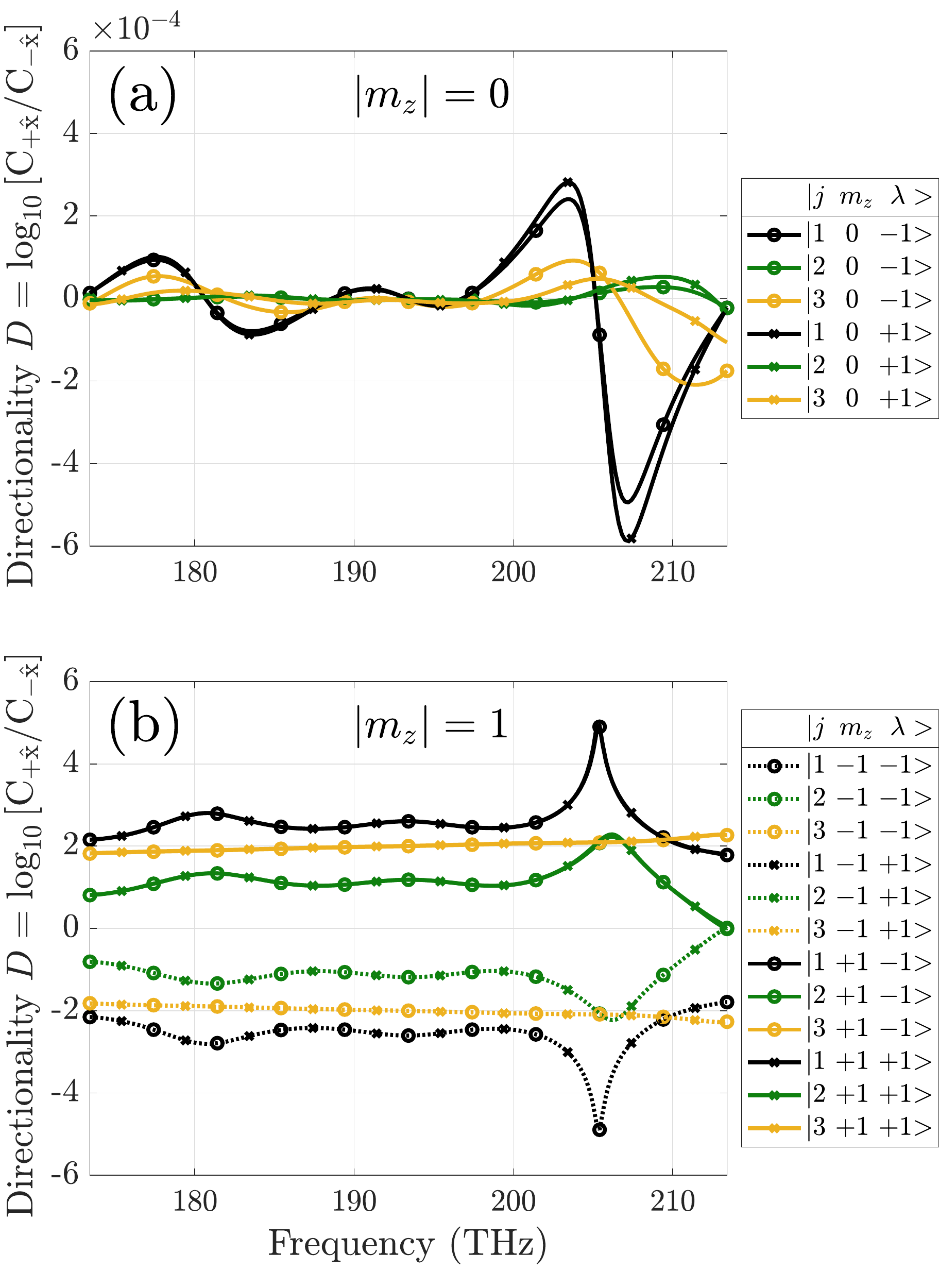}
	\caption{\label{fig:fig3} Directionality $D$ with respect to frequency for the coupling of multipolar emissions $\ket{\mathrm{k}\hspace{2pt}j\hspace{2pt}m_z\hspace{2pt}\lambda}$ with $j={1,2,3}$ and $m_z=0$ (a), or $|m_z|=1$ (b). The emitter is located in the xOy plane.}
\end{figure}

In our simulations, each emission contains a single helical multipole. The helical multipoles, or multipoles of well-defined helicity are linear combinations of the electric and magnetic multipoles (\cite[Eq.~(11.4-25)]{tung1985group}, \cite[Eq.~(2.18)]{Lakhtakia1994}). The salient characteristic of the helical multipoles is that all the plane-waves in their decomposition, including the evanescent ones, have the same polarization handedness. We denote the helical multipoles by $\ket{\mathrm{k}\hspace{2pt}j\hspace{2pt}m_z\hspace{2pt}\lambda}$, where $\mathrm{k}$ is the wavenumber, $j$ is the multipolar order (dipole $j=1$, quadrupole $j=2$, etc ...), $m_z\in[-j,-j+1,\ldots,j-1, j]$ is the projection of angular momentum along the $z$ axis, and $\lambda\in\{-1,+1\}$ is the helicity or handedness. Appendix~\ref{app:math} contains explicit expressions and relevant properties of helical multipoles. Any emission can be decomposed into helical multipoles. They form a complete basis for the fields radiated by an arbitrary emitter. For example, the fields emitted by an arbitrarily-oriented electric dipole $\dd$ of frequency $\omega=\mathrm{k}c_0$ can be written as (see App.~\ref{app:dipoles} ):
\begin{equation}
	\label{eq:dipole}
	\sum_{m_z=-1}^{m_z=1} p_{m_z} \left(|\mathrm{k}\ 1\ m_z\ +\rangle-|\mathrm{k}\ 1\ m_z\ -\rangle\right),
\end{equation}
where the weights $p_{m_z}$ are proportional to the projection of the spherical basis vectors $\{\hat{\mathbf{e}}_{-1}=(\xhat-\ii\yhat)/\sqrt{2},\hat{\mathbf{e}}_0=\zhat,\hat{\mathbf{e}}_1=-(\xhat+\ii\yhat)/\sqrt{2}\}$ onto $\dd$.

We consider emissions up to the octupolar order ($j=3$) for both helicities and all possible values of $m_z$, for a total of 30=(3+5+7)$\times$2 cases. This allows us to study the separate effect that angular momentum and helicity\footnote{While angular momentum and handedness can coincide for some particular fields, like circularly polarized Gaussian beams or single plane-waves, this is not true in general. For example, the handedness of vortex beams can be chosen independently from their angular momentum \cite{Zambrana2016}. Angular momentum and helicity are represented by two different commuting operators which generate two distinct symmetry transformations\cite{FerCor2012b,FerCor2012p}: Angular momentum generates rotations and helicity generates electromagnetic duality\cite{Calkin1965,Zwanziger1968}, whose action in momentum space is a rotation of the polarization of each plane-wave.}, i.e. the rotational or chiral properties of the emissions may have on the coupling directionality. We note that angular momentum and handedness can be most easily confused in the dipolar approximation. The field radiated by what is commonly referred to as\cite{LeFeber2015,Sollner2015,Coles2016,Picardi2017} {\em circular-dipole} or {\em circularly-polarized electric dipole moment} $\dd=-\xhat-\ii\yhat(\dd=\xhat-\ii\yhat)$, has a single non-zero coefficient $p_{1}(p_{-1})$ in \Eq{eq:dipole}. The radiation of a {\em circularly-polarized electric dipole} has hence a well-defined angular momentum $m_z=1$ or $m_z=-1$, but is a perfect mix of the two handedness in both cases. When our results are particularized to the dipolar approximation, it is seen that the directional coupling effect is controlled by the $\pm$ sign in $\dd=\pm\xhat- \ii\yhat$ {\em because such sign determines the transverse angular momentum} $m_z=\mp 1$ of the radiated fields, not because it reflects the handedness of the emitted light. Crucially, while such $\pm$ sign is reminiscent of a binary property like chirality or handedness, it should not be identified with it. Such incorrect identification suggests that the effect is always binary, while, as we show in this article, it rather features an $m_z$-dependent non-binary gradation. 
 
In our simulations, the directionality is computed as follows. After the emission, a portion of the radiated power couples into the waveguide. The power coupled to the first waveguide mode travelling towards either the $+\hat{\mathrm{x}}$ or the $-\hat{\mathrm{x}}$ direction is recorded by two dedicated ports. We refer to the power coupled towards the $\pm\hat{\mathrm{x}}$ direction as $\mathrm{C}_{\pm\hat{\mathrm{x}}}$. Figure~\ref{fig:fig2} shows the logarithmic directionality of the in-coupled power $D=\mathrm{log}_{10}\left[\mathrm{C}_{+\hat{\mathrm{x}}}/\mathrm{C}_{-\hat{\mathrm{x}}}\right]$ for varying angular momentum $(m_z,j)$ and helicity $\lambda$. A positive(negative) $D$ indicates preferential coupling towards the $+\hat{\mathrm{x}}(-\hat{\mathrm{x}})$ direction, and $|D|$ measures the degree of directionality in a logarithmic scale. For each $(m_z,j)$, the data in blue(red) corresponds to the positive(negative) helicity. The color intensity encodes the frequency distribution of $D$ as indicated by the insets. On the one hand, Fig.~\ref{fig:fig2} clearly shows that the helicity does not influence the value of $D$: Emissions with the same multipolar content $(m_z,j)$ but opposite helicity produce the same values of $D$\footnote{The small discrepancies for large $|m_z|$ are due to the low signal to noise ratio of the simulation results in the non-preferred direction.}. We will later show that this follows from the symmetries of the system. On the other hand, Fig.~\ref{fig:fig2} shows a clear dependence of $D$ on the transverse angular momentum $m_z$, which approximately follows\footnote{The exact slope depends in the particularities of the system, as discussed in App.~\ref{app:PWS}.} the green dashed line corresponding to $2m_z$. The sign of $m_z$ fixes the preferential coupling direction and, in a linear scale, the degree of directionality grows exponentially as $\approx 2|m_z|$. We observe that two emissions a and b, with $(m_z,j)_{\text{a}}$ and $(-m_z,j)_{\text{b}}$ result in $D_{\text{a}}=-D_{\text{b}}$, and that for $m_z=0$, $D\approx 0$. These regularities will be also shown to follow from the symmetries of the system. Figure~\ref{fig:fig3} shows $D$ as a function of frequency for some exemplary cases.

All the regularities are clearly visible across the whole spectrum. In particular, we observe in Fig.~\ref{fig:fig3}(a) that there is essentially no preferential coupling direction when $m_z=0$ (note the vertical scale). Half the in-coupled power travels towards each direction. The small fluctuations around $D=0$ can be attributed to numerical errors. Figure~\ref{fig:fig3}(b) shows that the directionality of all multipolar emissions with $m_z=+1$($m_z=-1$) is positive(negative). In Fig.~\ref{fig:fig3}(b) we clearly observe that the directionality of a particular $(m_z,j)$ emission is opposite to the directionality of the $(-m_z,j)$ emission, and that there is a perfect spectral overlap of the directionality of multipolar emissions with equal $(m_z,j)$ but opposite helicities. In particular, these results demonstrate that the orientation of a dipolar emitter, being directly related with the transverse angular momentum of the emission via \Eq{eq:dipole}, determines the directional coupling, while the helicity of the emission has no influence on it.

\begin{figure}
\includegraphics[width=8.45cm]{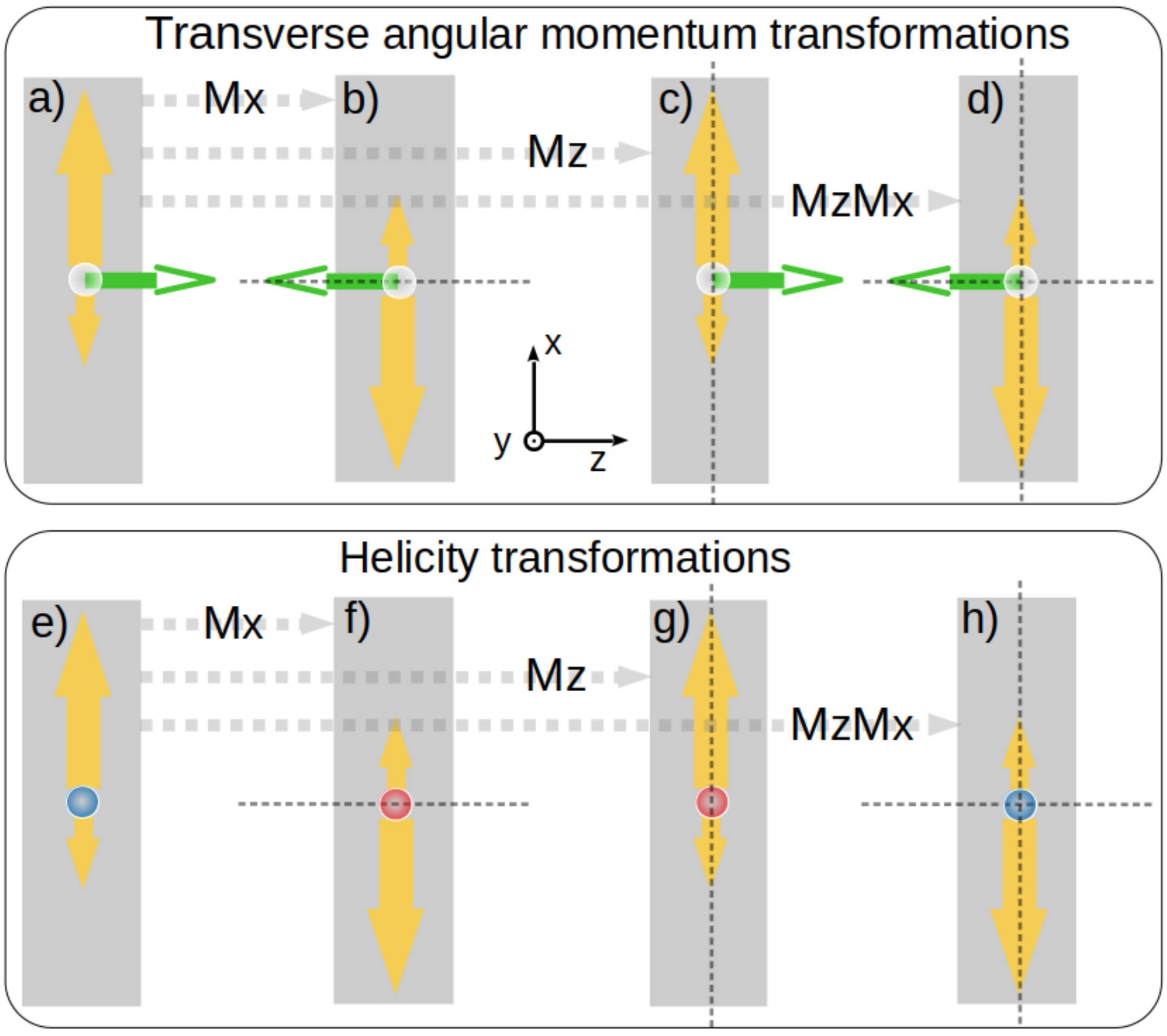}
	 \caption{Transformations of transverse angular momentum (green arrows), helicity (red/blue circles), and in-coupled power flux (yellow arrows) upon different reflection symmetries of the waveguide (gray strips). The initial situations [panels a) and e)] are transformed by $\Mx$ [panels b) and f)], $\Mz$ [panels c) and g)], and the composition $\Mz\Mx$ [panels d) and h)], respectively. In each panel, the origin of coordinates is at the position of the emitter, and the coordinate axes are oriented as shown in the figure.\label{fig:transformations}}
\end{figure}

\section{Symmetry analysis}
We now use the invariance of the waveguide upon reflection across the planes yOz ($\Mx$) and xOy ($\Mz$) to infer several fundamental characteristics of the directional coupling effect from the transformations of the joint emission-waveguide system. The mirror reflection properties of the helicity ($\Lambda=\mathbf{J}\cdot\mathbf{P}/|\mathbf{P}|$) and angular momentum ($\JJ=m\zhat$) of the emissions, and of the power flow towards the $\pm\xhat$ directions inside the waveguide ($\FF=F\xhat$) are hence of relevance. Such transformation properties are readily derived\footnote{We use the following decomposition of a reflection across a plane perpendicular to a unit vector $\alphahat$: $\mathrm{M}_\alpha=R_\alpha(\pi)\Pi$, where $\Pi$ is the parity transformation and $R_\alpha(\pi)$ a rotation by $\pi$ along $\alphahat$. Under rotations, angular momentum ($\JJ$) and linear momentum ($\PP$) or the Poynting vector behave as vectors, while helicity ($\Lambda$) is invariant. Under parity, helicity is a pseudo-scalar which changes sign $\Pi\left(\Lambda\right)\rightarrow -\Lambda$, while angular momentum $\JJ$ and linear momentum $\mathbf{P}$ exhibit the different parity transformation properties of axial and polar vectors, respectively: $\Pi\left(\JJ\right)\rightarrow\JJ$, versus $\Pi\left(\PP\right)\rightarrow-\PP$.} by noting that the properties of the power flow must be akin to those of linear momentum and the Poynting vector, and that angular momentum and linear momentum transform differently under parity and mirror symmetries due to their axial and polar vector character, respectively: 
	\begin{eqnarray}\nonumber
		&&\Mx\left(\JJ\right)=\Mx\left(m\zhat\right)=R_x(\pi)\left[\Pi\left(m\zhat\right)\right]=R_x(\pi)\left(m\zhat\right)\rightarrow -m\zhat,\\\nonumber
		&&\Mz\left(\JJ\right)=\Mz\left(m\zhat\right)=R_z(\pi)\left[\Pi\left(m\zhat\right)\right]=R_z(\pi)\left(m\zhat\right)\rightarrow m\zhat,\\\nonumber
		&&\Mx\left(\FF\right)=\Mx\left(F\xhat\right)=R_x(\pi)\left[\Pi\left(F\xhat\right)\right]=R_x(\pi)\left(-F\xhat\right)\rightarrow -F\xhat,\\\nonumber
		&&\Mz\left(\FF\right)=\Mz\left(F\xhat\right)=R_z(\pi)\left[\Pi\left(F\xhat\right)\right]=R_z(\pi)\left(-F\xhat\right)\rightarrow F\xhat,\\
		&&\Mx\left(\Lambda\right)\rightarrow -\Lambda,\ \Mz\left(\Lambda\right)\rightarrow -\Lambda.\label{eq:transformations}
	\end{eqnarray}

Figure~\ref{fig:transformations} shows the transformations of the initial situations [panels a) and e)], upon the following symmetries of the waveguide: $\Mx$ [panels b) and f)], $\Mz$ [panels c) and g)], and the composition $\Mz\Mx$ [panels d) and h)]. Angular momentum is represented by green arrows, positive(negative) helicity by blue(red) circles, and power flux by yellow arrows of different size reflecting a preferred coupling direction. The angular momentum and helicity of the emission are separately considered in panels a) to d), and e) to h), respectively. In the initial, yet untransformed, situation of panels a) and e) we {\em hypothesize} some degree of directional coupling depending solely on angular momentum and solely on helicity, respectively. Such hypothesis is falsified when a transformed system shows a physical contradiction with the original one. For example, the emission in panel g) occurs from the same location as the emission in panel e), and, even though the emissions have opposite helicity, they result in the same directionality. Similarly, the emission of panel h) is from the same location and of the same helicity as in panel e), but results in the opposite directionality. No such contradictions can be found regarding angular momentum dependent directionality when $|m_z|>0$. The comparison of panel a) with panels b,c,d) only shows that the directionality changes sign with the sign of the transverse angular momentum. The case $m_z = 0$ is special because it is invariant under the action of $\Mx$: $m_z\rightarrow-m_z$ [\Eq{eq:transformations}]. This leads to a contradiction between panels a) and b), where the same value of $m_z$ results in opposite directionality. Algebraic derivations can be found in App.~\ref{app:dem}, where we show that the $\Mz$ symmetry implies that two helical emissions $|k\ j\ m_z\ \lambda\rangle$ and $|k\ j\ m_z\ -\lambda\rangle$ will have the same directionality, and that the $\Mz\Mx$ symmetry implies that $|k\ j\ m_z\ \lambda\rangle$ and $|k\ j\ -m_z\ \lambda\rangle$ will have opposite directionality. The latter implies $D=0$ for $m_z=0$. The simulation results obey all these symmetry-based predictions. The same regularities will occur in any other geometry with the same symmetries. For example: i) The same system as in Fig.~\ref{fig:fig1} but with the waveguide turned 90 degrees along its axis; ii) The same system as in Fig.~\ref{fig:fig1} or i) but with a substrate parallel to the xOz plane supporting the waveguide; and iii) A cylindrical waveguide or a tapered fiber. 

Let us now analyze the directionality for centered electric and magnetic multipolar emissions with fixed ($\mathrm{k},j,m_z$). We derive the following selection rule in App.~\ref{app:elmagdir}: An electric or magnetic multipolar emission can only couple to a waveguide mode when $\tau q(-1)^{j+m_z}=1$, where $\tau=+1(-1)$ for electric(magnetic) multipoles and $q$ is the $\Mz$ eigenvalue of the waveguide mode. We note that two counter-propagating but otherwise identical modes have the same value of $q$. Therefore, when the selection rule is not met, the emitter cannot couple to any of the two counter-propagating modes, and when the selection rule is met, the emitter can couple to both modes. When the coupling is allowed, the directionality for the electric(magnetic) multipoles will be the same as the directionality for the helicity multipoles (see App.~\ref{app:elmagdir}). In fact, the directionality will be the same for any linear combination of the two helical multipoles $|k\ j\ m_z\ \ \lambda=\pm\rangle$. 

Figure.~\ref{fig:transformations} helps elucidating other properties of the directional coupling effect. 

\begin{figure}[h!]
\includegraphics[width=8.45cm]{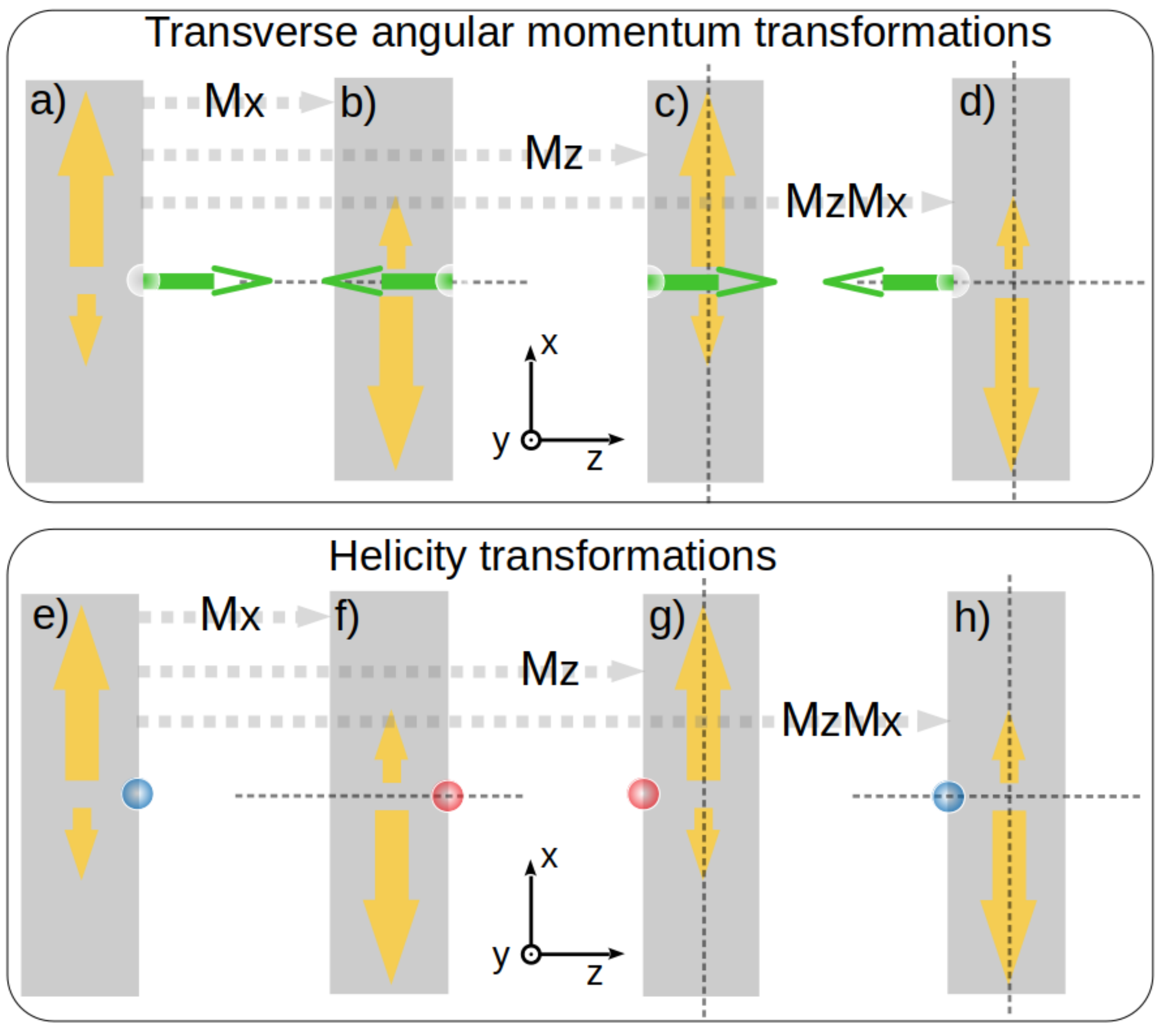}
		\caption{\label{fig:edge} Transformations of transverse angular momentum (green arrows), helicity (red/blue circles), and in-coupled power flux (yellow arrows) upon different reflection symmetries of the waveguide (gray strips) for an emitter displaced by $d_z=100$nm from the xOy plane. In each panel, the origin of coordinates is at the position were the emitter was before the displacement (see Figs.~\ref{fig:fig1},\ref{fig:transformations}) and the coordinate axes are oriented as shown in the figure.}
\end{figure}

The directionality changes sign upon $\Mx$ [panels b) and f)]. This implies the intuitive fact that the emission must break the $\Mx$ symmetry in order for it to couple directionally. Otherwise, invariance of the emission combined with the change of sign of the directionality would imply $D=0$. This necessary condition is met by both transverse angular momentum and helicity, which change upon $\Mx$. In light of this, a helicity-dependent directionality may be possible for an emitter displaced out of the xOy plane. The reason is that the displacement breaks the $\Mz$ reflection symmetry that forbids helicity-dependent directionality for in-plane emitters. This is illustrated in Fig.~\ref{fig:edge}: The contradictions between Fig.~\ref{fig:transformations}(e) and Fig.~\ref{fig:transformations}(g,h) do not occur between Fig.~\ref{fig:edge}(e) and Fig.~\ref{fig:edge}(g,h) because the emitter is not at the original position. 

\begin{figure}[h!]
\includegraphics[width=8.45cm]{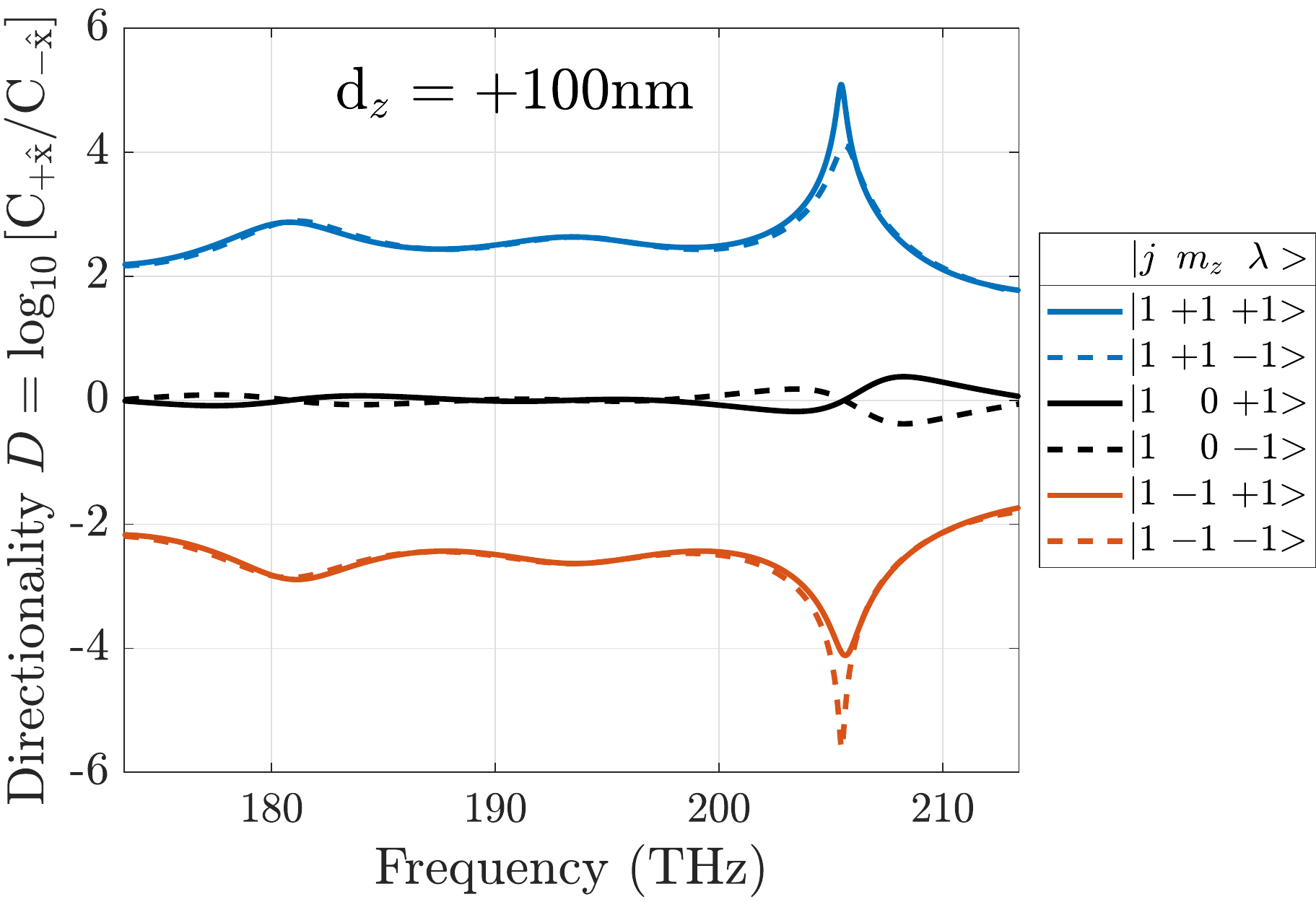}
	\caption{\label{fig:fig5} Directionality $D$ with respect to frequency for the coupling of multipolar emissions $\ket{\mathrm{k}\hspace{2pt}j\hspace{2pt}m_z\hspace{2pt}\lambda}$ with $j=1$, $m_z\in\{-1,0,1\}$, and $\lambda\in\{-1,1\}$. The emitter is displaced by $d_z=100$nm from the xOy plane, as shown in Fig.~\ref{fig:edge}(a).}
\end{figure}

We confirm the existence of position and helicity dependent directionality for displaced emitters by numerical simulations. Figure~\ref{fig:fig5} shows the directionality of dipolar emissions for an emitter that has been displaced out of the xOy plane by $100$nm along the positive $z$ direction, right to the vertical of the edge of the waveguide, as shown in Fig.~\ref{fig:edge}(a). Helicity has now some influence on directionality. For example, some directionality can be observed in Fig.~\ref{fig:fig5} for $m_z=0$, in contrast to the centered case in Fig.~\ref{fig:fig3}(a). Also, when $|m_z|=1$ the curves for $|j=1\ m_z\ \lambda\rangle$ and $|j=1\ m_z\ -\lambda\rangle$ are not on top of each other, as is the case for the centered emitter in Fig.~\ref{fig:fig3}(b). Nevertheless, the sign of $m_z=\pm 1$ still determines the sign of $D$. The comparison between Figs.~\ref{fig:fig3} and \ref{fig:fig5} shows that the influence of helicity in the directionality of a displaced emitter is rather small \footnote{The largest influence can be observed at the sharp spectral features in the case of dipolar emissions at 205.3~THz. The features are due to a pronounced dip in the frequency-dependent coupling into the guided mode towards the non-preferred direction, leading to a spectral peak of the directionality towards the opposite direction.} when compared to the influence of angular-momentum when $|m_z|>0$. Figure~\ref{fig:fig5} also shows that the coupling direction favored by a given value of helicity changes with frequency. This, and the fact that the same helicity will produce the opposite directionality depending on its position [see Figs.~\ref{fig:edge}(e,h)], are in sharp contrast with the effect of transverse angular momentum, which is very similar for the cases of centered and displaced emitters. Such similarity is not unexpected after considering the top parts of Figs.~\ref{fig:edge} and \ref{fig:transformations}. The results indicate that the influence of transverse angular momentum in the directionality is largely independent of the lateral position of the emitter. We note that other geometries like cylindrical waveguides do not allow helicity-dependent directionality for emissions with well defined j and well defined angular momentum $m_\alpha$ with respect to an axis ${\mathbf{\hat{\alpha}}}$ transverse to the waveguide axis. This follows because then, $\mathrm{M}_{\mathrm{\alpha}}$ would play the role played previously by $\Mz$ in showing that the directionality could not depend on helicity. Finally, we observe that Fig.~\ref{fig:fig5} clearly shows the regularity due to the $\Mx$ symmetry of the waveguide: Independently of the position of the emitter, two helical emissions $|k\ j\ m_z\ \lambda\rangle$ and $|k\ j\ -m_z\ -\lambda\rangle$ will have the opposite directionality (App.~\ref{app:dem}).

From now on, we focus on the dominant directionality effect, where the emissions break the $\Mx$ symmetry due to the axial character of the transverse angular momentum [compare panels a) and b)]. The dominant directionality is hence due to an axial vector (transverse angular momentum), not to a pseudo-scalar (helicity). Interestingly, the symmetry breaking by axial vectors has been studied in the context of enantio-selective chemical reactions, where Barron refers to it as ``false chirality'' (see \cite{Barron2020} and the references therein). The correct identification of the origin of the directionality is crucial for understanding that it is not a binary effect: While a pseudo-scalar offers only two possibilities which can explain the sign of $D$, an axial vector can explain the sign and magnitude gradation of $D$ through the sign and magnitude of the vector, respectively. Since the $\Mx$ symmetry is broken by the $m\zhat \rightarrow -m\zhat$ change, the degree of $\Mx$ breaking must be related to the magnitude of the change ($|2m|$), which vanishes for $m=0$, suggesting that $D$ should grow with $|2m|$. We show in App.~\ref{app:PWS} that the growth is exponential.

Finally, Fig.~\ref{fig:transformations} also allows us to determine whether the directional coupling effect is chiral, as is often stated in the literature. Panels c) and g) show that the directionality is invariant upon a mirror reflection ($\Mz$) of the emission. The effect has hence a mirror symmetry, which makes it achiral\footnote{The definition of chirality is (see e.g. \cite[Chap. 2.6]{Bishop1993}) the lack of: Parity, all mirror reflections, and all improper rotation symmetries.}.

\section{Exponential directionality} 
The exponential dependence of the directionality on the transverse angular momentum is remarkable. Its origin can be traced back to an intrinsic property of the evanescent angular spectrum of the multipolar emissions. Appendix~\ref{app:PWS} shows that: i) Only the evanescent plane-waves in the angular spectrum of $\ket{\mathrm{k}\hspace{2pt}j\hspace{2pt}m_z\hspace{2pt}\lambda}$ can couple power into the waveguide, and ii) The power flux (real part of the Poynting vector) carried by those evanescent plane-wave components towards the $\pm \hat{\mathrm{x}}$ directions is proportional to a term that has a $\pm m_z$ exponential dependence. The origin of the exponential directionality is hence an intrinsic property of the emissions, independent of the details of the waveguide. This generality is consistent with the wide variety of experimental setups where the directional coupling due to transverse angular momentum has been observed \cite{Mitsch2014b,Petersen2014,Rodriguez2014}. The exponential directionality is also in particular consistent with Ref.~\onlinecite{transversecouplingevanescentwaves}, where the transverse angular momentum content of evanescent plane-waves was shown to also depend exponentially on the eigenvalue of transverse angular momentum\footnote{Denoted by $m_y$ in that work due to a different axis orientation. The fact that the result in Ref.~\onlinecite{transversecouplingevanescentwaves} can be used to explain the way atomic transitions are excited by the evanescent tails of guided modes \cite{Mitsch2014,Le2014} strongly suggests that the directionality of the coupling of an emitter into guided modes should also depend exponentially on the transverse angular momentum: The excitation of a guided mode by an emitting object can be seen as the reciprocal situation from the one where the object is excited by the guided mode.}.

\section{A different angular momentum axis}
The axis of well-defined angular momentum for the emission has been experimentally selected by i) the direction of a static magnetic field, or ii) the axis along which the incident excitation has a well-defined angular momentum. When a static magnetic field is used, the frequency of the emissions with different angular momenta will be spectrally split by the Zeeman effect. In some experiments the choice of axis has coincided with what we define as the transverse axis $\zhat$ \cite{Mitsch2014b,Petersen2014,Rodriguez2014}. In other experiments, a different axis has been chosen \cite{Sollner2015,Coles2016,LeFeber2015,Scarpelli2019}, coinciding with our $\yhat$ axis. 

\begin{figure}
\includegraphics[width=8.45cm]{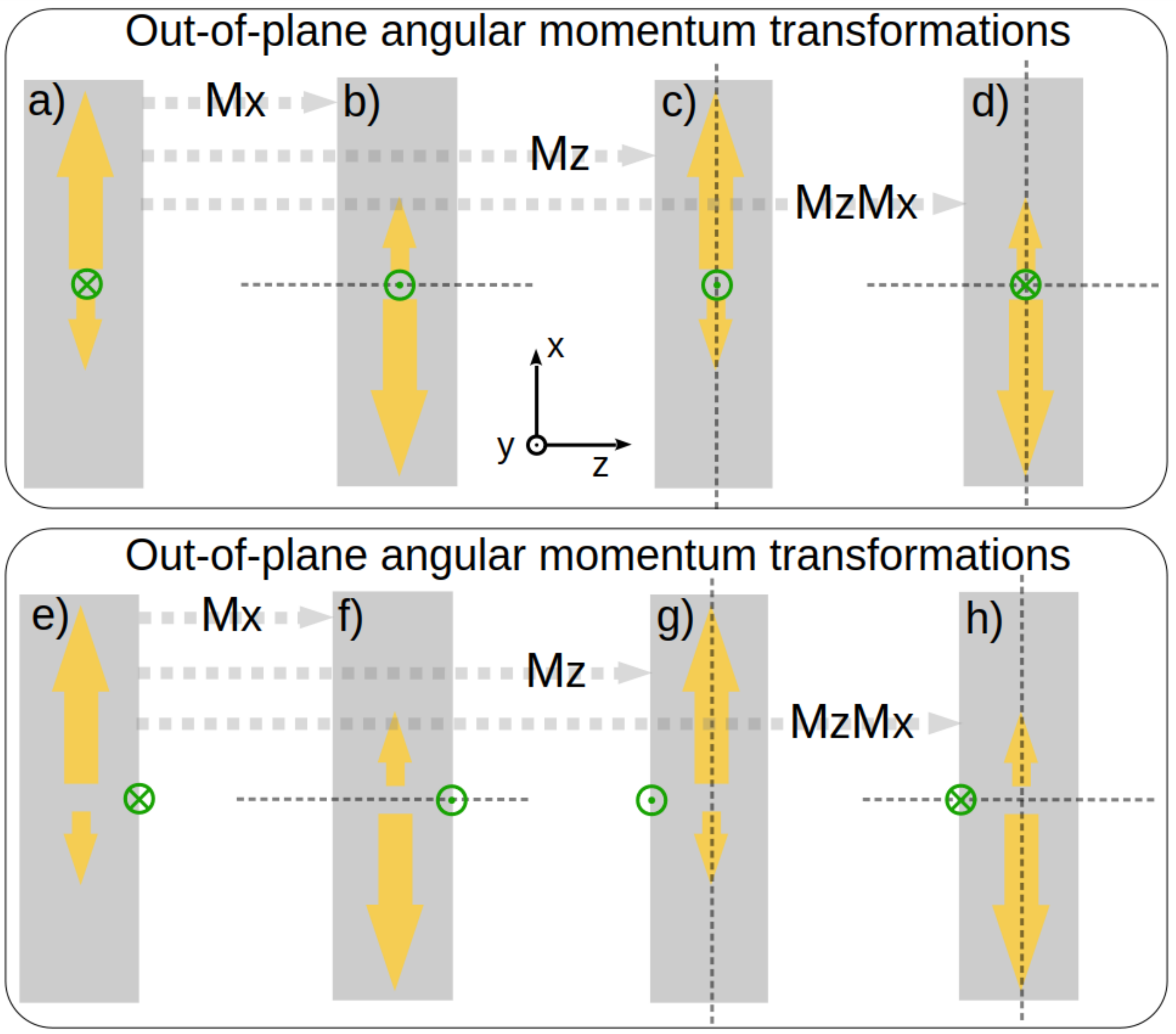}
		\caption{Transformations of the out-of-plane $m_y\yhat$ angular momentum (green symbols) and in-coupled power flux (yellow arrows) upon different reflection symmetries of the waveguide (gray strips). The top(bottom) part shows a centered(displaced) emitter.\label{fig:transformations_outofplane}}
\end{figure}

The behavior of the directionality $D$ with respect to $m_y$ and $m_z$ is different. In Fig.~\ref{fig:transformations_outofplane} we use the transformation properties $\Mx\left(m_y\yhat\right)\rightarrow-m_y\yhat$, and $\Mz\left(m_y\yhat\right)\rightarrow-m_y\yhat$ to analyze the cases of both centered and displaced emitters.
In the top part of the figure, a contradiction between panels a) and panels (c,d) is seen for a centered emitter. Such contradiction does not occur for the displaced emitter in the bottom part of the figure because the emitter is not in the original position. Appendix~\ref{app:jy} contains a proof that, in particular, shows that on a waveguide with both $\Mz$ and $\Mx$ symmetries the directionality for a centered emitter with well-defined $\yhat$ component of angular momentum will be identically zero. This explains the lack(existence) of directionality for centered(displaced) emitters measured in Refs.~\onlinecite{Coles2016,LeFeber2015}. Moreover, it also explains the possibility of directionality from centered emitters {when the waveguide itself lacks the $\Mz$ reflection symmetry}, as in Ref.~\onlinecite{Sollner2015}. In this case, the waveguide in panel a) would be different from those in panels (c,d), avoiding the contradiction.

We also deduce from the bottom part of Fig.~\ref{fig:transformations_outofplane} that, in sharp contrast to the $m_z$ case, the sign of the angular momentum dependent directionality {\em is not locked} to the sign of $m_y$. For example, panels e) and h) show the same sign of $m_y$ but opposite directionality. This dependence of the directionality on the lateral ($z$) position for fixed $m_y$ has been experimentally observed \cite[Fig.~3(b)]{LeFeber2015}. On the other hand, if the angular momentum axis is chosen along the transverse direction ($z$) instead of along the out-of-plane direction ($y$), then, the preferred coupling direction will be locked to the sign of the angular momentum of the emissions independently of the lateral position of the emitter\footnote{This statement can be inferred graphically from the top row of Fig.~\ref{fig:edge}, and Fig.~\ref{fig:fig5} provides a numerical example for it.}. This suggests that changing the magnetic bias direction in setups like those in Refs.~\cite{Sollner2015,Coles2016,Scarpelli2019} will potentially have a significant effect. Let us analyze this possibility. 

When considering such change, another difference between the experiments in Refs.~\cite{Mitsch2014b,Petersen2014,Rodriguez2014} and those in Refs.~\cite{Sollner2015,Coles2016,Scarpelli2019} becomes consequential: In the former experiments the emitters are on top of the waveguides, while in the latter experiments the layer of quantum dots is embedded inside the waveguide. Here, ``on top'' means that the emitters are above the waveguide as in Figs~.(\ref{fig:transformations},\ref{fig:edge},\ref{fig:transformations_outofplane}). While the symmetry arguments illustrated in Figs.~(\ref{fig:transformations},\ref{fig:edge},\ref{fig:transformations_outofplane}) apply independently of the vertical ($y$) position of the emitters, it can be shown that, for fixed angular momentum $m_z$, the preferred coupling direction changes sign when moving the emitter from the top to the bottom of the waveguide \cite{transversecouplingevanescentwaves}, and zero directionality is expected from quantum dots that are embedded exactly in the middle $y$-coordinate of the waveguide. 

Before continuing, it should also be noted that the experiments in Refs.~\onlinecite{Sollner2015,Coles2016,Scarpelli2019} address individual quantum dots out of the many quantum dots of the waveguide, which have roughly random $x$ and $z$ positions, and a fixed $y$ position. In these experiments, the pump illumination is focused onto a single quantum dot at a particularly favorable position regarding directionality, and such quantum dot is the only contributor to the signals detected at each end of the waveguide. The signals originating from $m_y=\pm 1$ transitions are distinguished at the detectors because they are produced at different frequencies due to the Zeeman splitting induced by the magnetic bias. 

We now propose an experiment where the layer of quantum dots is at (very near to) the top of the waveguide, the static magnetic field is applied along the $z$ direction and, instead of addressing a single quantum dot, the pump illuminates a much larger area, potentially including many more quantum dots. It is clear from previous discussions that all the $m_z=+1$ emitters will preferentially couple towards one direction and all the $m_z=-1$ emitters will preferentially couple to the opposite one. Therefore, besides showing directionality, the signals at the detectors should be significantly larger than the signals from an individual quantum dot. Assuming incoherent emissions, the total detected average power (photon count) will be larger than the detected average power (photon count) of individual quantum dots by a factor equal to the number of quantum dots excited by the pump. Moreover, a super-radiant scaling cannot be ruled out \cite{Kornovan2017}. Note that illuminating a larger area when the bias is along the $y$ direction leads to approximately zero directionality. This is clear from the facts that, for fixed $m_y$, the sign of the directionality changes when the emitter is in opposite sides of the waveguide (Fig.~\ref{fig:transformations_outofplane}), and that the $z$ positions of the quantum dots are random. 

We highlight that the aggregation of the directional signals of many quantum dots occurs for any multipolar order of the emissions, including the dipolar one. The same proposed experimental setup can be used to test other quantitative and qualitative predictions contained in this article if emitters with $|m_z|>1$ are available.

\section{Final remarks} 
In this article, we have identified the symmetry and symmetry-breaking mechanisms behind the directional coupling of emitters into nearby waveguides. We have also shown that the directionality is mostly determined by the transverse angular momentum, whose sign determines the preferential coupling direction, and whose absolute value affects the degree of directionality in an exponential way. We have made several predictions for yet unobserved effects, including an experiment featuring a transverse magnetic bias for the aggregation of the directional emissions of quantum dots on top of waveguides.

Regarding other plausible applications of the dominant directional coupling effect: On the one hand, the exponential $m_z$ dependence and the selection rule for electric and magnetic multipolar emissions may be exploited for routing, detecting, and classifying the transitions of discrete nano-emitters. In particular, our framework is specifically suited for understanding and predicting the directional coupling of higher-order multipolar transitions \cite{Stourm2020}, which are the object of an increasing number of experimental \cite{Tojo2004,Andersen2011,Cheng2012,Karaveli2010,Kasperczyk2015,Vaskin2019}, and theoretical studies \cite{Zurita2002,Kern2012,Dodson2012,Konzelmann2019,Stourm2020}. On the other hand, contrary to what is sometimes claimed \cite{Rodriguez2014,Fang2019}, $D$ does not allow to distinguish the helicity, chirality or handedness of the emission, and hence the consequently suggested applications for chiral molecule sensing \cite{Kapitanova2014} are not possible.

\begin{acknowledgments}
	We warmly thank Ms.~Maria Lamprianidou for drawing Fig.~\ref{fig:fig1} , and Ms.~Magda Felo for her help with Figs.~(\ref{fig:transformations},\ref{fig:edge},\ref{fig:transformations_outofplane}). We acknowledge financial support by the Max Planck School of Photonics, which is supported by BMBF, Max Planck Society, and Fraunhofer Society, and by the Deutsche Forschungsgemeinschaft (DFG, German Research Foundation) under Germany’s Excellence Strategy via the Excellence Cluster 3D Matter Made to Order (EXC-2082/1 – 390761711) through its CZF-Focus@HEiKA graduate school. X.Z.-P. acknowledges funding from the European Union's Horizon 2020 research and innovation programme under the Marie Sklodowska-Curie grant agreement No 795838.
\end{acknowledgments}

\ifdefstring{\includeappendix}{no}{\end{document}}{}
\ifdefstring{\includeappendix}{yes}{\appendix\setcounter{secnumdepth}{2}\renewcommand\thefigure{S\arabic{figure}}\section{Details about the numerical simulation with CST MWS\label{app:cst}}
Each emission is modeled with the help of an imaginary auxiliary box surrounding the emitter (see Fig.~\ref{fig:fig1}). For a given helical multipole $\ket{\mathrm{k}\hspace{2pt}j\hspace{2pt}m_z\hspace{2pt}\lambda}$, the exact radiated electric and magnetic fields on the surface of the box are computed using Eqs.~(\ref{eqnMultDef},\ref{eqnEMfields}). Then, their tangential components are imprinted on the surface of the box as electric and magnetic source currents. According to the surface equivalence principle (Chap. 3.5 in Ref.~\onlinecite{Harrington1961}), the electromagnetic field that these sources produce outside the box is identical to the electromagnetic field emanating from the multipolar emission from the center of the box. We pick the size of the auxiliary box to be 190~nm and assign a mesh step of 3.8~nm across its surface. Picking a fine mesh here is crucial because the evanescent fields generated by the multipolar emission need to be accurately modeled since they are the ones responsible for the near-field coupling to the waveguide (see App.~\ref{app:PWS}). A mesh step of $0.02\lambda_0$ was chosen, which allows us to correctly model fast varying evanescent fields with a spatial periodicity of even about $0.1\lambda_0$.

 Open boundary conditions are selected everywhere. The waveguide ports that collect the power guided from the emitter to each side of the waveguide are placed at a distance of \SI{6}{\micro\meter} from the yOz plane.

Finally, we note that all our formulas have an implicit $e^{-\mathrm{i}\omega t}$ time dependency, whereas CST MWS adopts the opposite convention $e^{\mathrm{i}\omega t}$. We therefore need to take special care to give the correct real fields. Specifically, we need to feed CST MWS with the complex conjugates of the formulas in Eqs.~(\ref{eqnMultDef},\ref{eqnEMfields}) so that $\mathrm{Re}\left\{\mathbf{E}^*_{\lambda,m_z j}(\mathrm{k}\mathbf{r})e^{\mathrm{i}\omega t}\right\}=\mathrm{Re}\left\{\mathbf{E}_{\lambda,m_z j}(\mathrm{k}\mathbf{r})e^{-\mathrm{i}\omega t}\right\}$ and $\mathrm{Re}\left\{\mathbf{H}^*_{\lambda,m_z j}(\mathrm{k}\mathbf{r})e^{\mathrm{i}\omega t}\right\}=\mathrm{Re}\left\{\mathbf{H}_{\lambda,m_z j}(\mathrm{k}\mathbf{r})e^{-\mathrm{i}\omega t}\right\}$.

\section{Multipoles of well-defined helicity\label{app:math}}
The multipoles of well-defined helicity $\mplambda$ that we use can be written as linear combinations of the electric and magnetic multipoles $\mptau$:
\begin{equation}
	\label{eqnheldef1}
	\mplambda=\frac{\mptauminus+\lambda\mptauplus}{\sqrt{2}},
\end{equation}
where $\tau=1(\tau=-1)$ corresponds to the electric(magnetic) multipoles, $\mathrm{k}$ is the wavenumber, $j=1,2,3,\ldots$ is related to $j(j+1)$, the eigenvalue of the total angular momentum squared operator $\mathbf{J}\cdot\mathbf{J}=J^2$, and $m_z\in[-j,-j+1,\ldots,j-1, j]$ is the eigenvalue of the $z$-component of the angular momentum $J_z$. We note that the different definition $\sqrt{2}\mplambda=\mptauplus+\tau\mptauminus$ is also possible. Both conventions are used in the literature (see e.g. \cite[Eq.~(11.4-19)]{tung1985group} versus \cite[Eq.~(2.18)]{Lakhtakia1994}). The $\mptau$ are eigenstates of the parity operator: $\Pi \ket{\mathrm{k}\hspace{2pt}j\hspace{2pt}m_z\hspace{2pt}\tau}=\tau(-1)^{j}\ket{\mathrm{k}\hspace{2pt}j\hspace{2pt}m_z\hspace{2pt}\tau}$.

The derivation of the $\rr$-dependent expressions of the $\mptau$ multipoles can be found in the literature (e.g. \cite[App. C]{Mishchenko2002}). Different conventions are again used by different authors, which then lead to different expressions for the radiating $\mplambda$ multipoles. We use the following one:

\begin{eqnarray}
	&&\ket{\mathrm{k}\hspace{2pt}j\hspace{2pt}m_z\hspace{2pt}\lambda} \equiv \frac{\lambda}{\sqrt{2}}\frac{j(j+1)}{\mathrm{kr}}h^{(1)}_j(\mathrm{kr})\gamma_j^{m_z}\mathrm{P}_j^{m_z}(\mathrm{cos}\theta)e^{\mathrm{i}m_z\phi} \hspace{3pt}\rhat\label{eqnMultDef}\\
&&+\frac{\mathrm{1}}{2}\left[\frac{1}{\mathrm{kr}}\frac{\mathrm{d}}{\mathrm{dkr}}\left(\mathrm{kr}h^{(1)}_j(\mathrm{kr})\right)+\mathrm{i}h^{(1)}_j(\mathrm{kr})\right]\mathrm{A}_{\lambda,m_z j}(\hat{\mathrm{r}}) \hspace{3pt}\hathellambdar\nonumber\\
&&-\frac{\mathrm{1}}{2}\left[\frac{1}{\mathrm{kr}}\frac{\mathrm{d}}{\mathrm{dkr}}\left(\mathrm{kr}h^{(1)}_j(\mathrm{kr})\right)-\mathrm{i}h^{(1)}_j(\mathrm{kr})\right]\mathrm{A}_{-\lambda,m_z j}(\hat{\mathrm{r}}) \hspace{3pt}\hathelminuslambdar,\nonumber
\end{eqnarray}
where $\mathrm{r}=|\rr|$, $h^{(1)}_j(\cdot)$ are spherical Hankel functions of the first kind, $\theta=\arccos(z/|\rr|)$, $\phi=\arctan2(y,x)$, $\gamma_j^{m_z}=\mathrm{i}^{m_z}\sqrt{(2j+1)(j-m_z)!}/\sqrt{4\pi j(j+1)(j+m_z)!}$, $\mathrm{P}_j^{m_z}(\cdot)$ are the associated Legendre function of the first kind, 
	{\small
\begin{eqnarray}
	\mathrm{A}_{\lambda,m_z j}(\hat{\mathrm{r}})&=& \gamma_j^{m_z}\left[-\frac{\mathrm{d}\mathrm{P}_j^{m_z}(\mathrm{cos}\theta)}{\mathrm{d}\theta}-\lambda m_z \frac{\mathrm{P}_j^{m_z}(\mathrm{cos}\theta)}{\mathrm{sin}\theta}\right]e^{\mathrm{i}m_z\phi},\label{eqnspectrum}\\
	\hathellambdar&=&\frac{-\lambda\thetar-\mathrm{i}\phir}{\sqrt{2}},\label{eqnpolvector}
\end{eqnarray}
	}
\noindent and $\{\rhat=\rr/\mathrm{r}, \thetar,\phir\}$ are the radial, polar, and azimuthal unit vectors that correspond to $\rr$. 

The electromagnetic field radiated by a particular multipolar emission of well-defined helicity is then:
\begin{equation}
	\mathbf{E}_{\lambda,m_z j}(\mathrm{k}\mathbf{r})\equiv\ket{\mathrm{k}\hspace{2pt}j\hspace{2pt}m_z\hspace{2pt}\lambda}, \hspace{15pt}\mathbf{H}_{\lambda,m_z j}(\mathrm{k}\mathbf{r})=\frac{\lambda}{\mathrm{iZ}}\mathbf{E}_{\lambda,m_z j}(\mathrm{k}\mathbf{r}),\label{eqnEMfields}
\end{equation}
where the rightmost expression follows from applying the Maxwell-Faraday equation $\mathbf{H}=\frac{\nabla\times}{kiZ}\mathbf{E}$ to fields of well defined helicity: Since $\frac{\nabla\times}{\mathrm{k}}$ is the representation of the helicity operator $\Lambda$ for monochromatic fields, then $\mathbf{H}=\Lambda\mathbf{E}/{\ii Z}=\lambda\mathbf{E}/{\ii Z}$.  

The transformation properties of the multipoles of well-defined helicity under mirror reflections and parity can be obtained from Eq.~(\ref{eqnMultDef}):

\begin{eqnarray}
	\label{eq:b6}
    \mathrm{M}_x\ket{\mathrm{k}\hspace{2pt}j\hspace{2pt}m_z\hspace{2pt}\lambda} &=&(-1)^{m_z+1}\ket{\mathrm{k}\hspace{2pt}j\hspace{2pt}-m_z\hspace{2pt}-\lambda},\label{eqnhelsym1a}\\
	\label{eq:b7}
\mathrm{M}_y\ket{\mathrm{k}\hspace{2pt}j\hspace{2pt}m_z\hspace{2pt}\lambda} &=&-\ket{\mathrm{k}\hspace{2pt}j\hspace{2pt}-m_z\hspace{2pt}-\lambda},\\
\mathrm{M}_z\ket{\mathrm{k}\hspace{2pt}j\hspace{2pt}m_z\hspace{2pt}\lambda} &=&(-1)^{j+m_z+1}\ket{\mathrm{k}\hspace{2pt}j\hspace{2pt}m_z\hspace{2pt}-\lambda},\label{eqnhelsym2a}\\
\Pi\ket{\mathrm{k}\hspace{2pt}j\hspace{2pt}m_z\hspace{2pt}\lambda}&=&(-1)^{j+1}\ket{\mathrm{k}\hspace{2pt}j\hspace{2pt}m_z\hspace{2pt}-\lambda}.
\label{eqnparityhel}
\end{eqnarray}

\section{Electric dipole radiation as superposition of helical multipoles of the two helicities\label{app:dipoles}}
A general radiation of frequency $\omega=\mathrm{k}c$ can be written as \cite[Chap. 9]{Jackson1998} and \cite[App. B, \S4]{Blatt1952}:
\begin{equation}
	\sum_{j=1}^\infty \sum_{m_z=-j}^{j} a^f_{jm_z}\mptauplus+b^f_{jm_z}\mptauminus,
\end{equation}
where the $\{a^f_{jm_z},b^f_{jm_z}\}$ are complex coefficients, namely the coefficients of the multipolar decomposition of the fields. The field radiated by an electric dipole can hence be written as:
\begin{equation}
	\begin{split}
		&a^f_{1,1}\ket{\mathrm{k}\hspace{2pt}1\hspace{2pt}1\hspace{2pt}\tau=1}+a^f_{1,0}\ket{\mathrm{k}\hspace{2pt}1\hspace{2pt}0\hspace{2pt}\tau=1}+a^f_{1,-1}\ket{\mathrm{k}\hspace{2pt}1\hspace{2pt}-1\hspace{2pt}\tau=1}=\\
		&\sum_{m_z=-1}^{1}a^f_{1,m_z}\ket{\mathrm{k}\hspace{2pt}1\hspace{2pt}m_z\hspace{2pt}\tau=1}\duetoeq{eq:taump}\\
		&\sum_{m_z=-1}^{1}\frac{a^f_{1,m_z}}{\sqrt{2}}\left(\ket{\mathrm{k}\hspace{2pt}1\hspace{2pt}m_z\hspace{2pt}\lambda=+1}-\ket{\mathrm{k}\hspace{2pt}1\hspace{2pt}m_z\hspace{2pt}\lambda=-1}\right),
	\end{split}
\end{equation}
which is a perfect mix of both helicities.

The $a^f_{1,m_z}$ are determined by the Cartesian components of an arbitrary electric dipole $\dd=p_x\xhat+p_y\yhat+p_z\zhat$. To see this, we first note that the $\{a^f_{jm_z},b^f_{jm_z}\}$ field coefficients must be proportional to the multipolar components of the emitting source $\{a^s_{jm},b^s_{jm}\}$. For an electric dipole the $a^s_{1,m_z}$ are also just proportional to the components of $\dd$ in the basis of spherical vectors $\{\hat{\mathbf{e}}_1=-(\xhat+\ii\yhat)/\sqrt{2},\hat{\mathbf{e}}_0=\zhat,\hat{\mathbf{e}}_{-1}=(\xhat-\ii\yhat)/\sqrt{2}\}$, namely \cite[Eq.~(C3)]{Alaee2016b}:
\begin{equation}
	a^s_{1,m_z}=\frac{\ii\omega p_{m_z}}{\sqrt{3}\pi},
\end{equation}
and
\begin{equation}
	\begin{bmatrix}p_{1}\\p_{0}\\p_{-1}\end{bmatrix}=\begin{bmatrix}\frac{-1}{\sqrt{2}}&\frac{i}{\sqrt{2}}&0\\0&0&1\\\frac{1}{\sqrt{2}}&\frac{\ii}{\sqrt{2}}&0\end{bmatrix}\begin{bmatrix}p_{x}\\p_{y}\\p_{z}\end{bmatrix}.
\end{equation}

\section{Symmetry derivations\label{app:dem}} 
In this Section we will demonstrate that the regularities observed in the numerical results shown in Fig.~\ref{fig:fig2} follow from symmetry arguments. Namely, we will show that:
\begin{enumerate}
	\item The directionality must be identical for multipolar emissions with equal $(\mathrm{k}, j, m_z)$ and opposite helicity $\lambda$, 
	\item the directionality must be opposite for multipolar emissions with equal $(\mathrm{k},j,\lambda)$ and opposite angular momentum $m_z$, and that
	\item the directionality must be zero for $m_z=0$.
\end{enumerate}
We will show that: i) The first statement follows from the mirror symmetry of the waveguide across the xOy plane, $\mathrm{M}_z$; ii) The second statement follows from the first, plus the mirror symmetry of the waveguide across the yOz plane, $\mathrm{M}_x$, and; iii) The third statement follows from the second one when $m_z=0$.

We model the coupling between the emission of a multipole $\ket{\mathrm{k}\hspace{2pt}j\hspace{2pt}m_z\hspace{2pt}\lambda}$ and the power guided along the $\pm\hat{\mathrm{x}}$ direction of the waveguide as:

\begin{equation}
\restr{\mathrm{C}_{\pm\hat{\mathrm{x}}}}{(\mathrm{k}, j, m_z, \lambda)} = \vert \bra{\sigma_{\pm \hat{\mathrm{x}}}} \hspace{2pt} \mathrm{S} \hspace{2pt} \ket{\mathrm{k}\hspace{2pt}j\hspace{2pt}m_z\hspace{2pt}\lambda} \vert^2
\label{eq:def_coupl}
\end{equation}

\noindent where $\sigma_{\pm \hat{\mathrm{x}}}$ is a guided mode of the waveguide in the $\pm \hat{\mathrm{x}}$ direction, and $\mathrm{S}$ is the S-matrix of the system that includes the coupling mechanism.
The directionality $D$ is defined as:

\begin{eqnarray}
	\label{eq:ddef}
    \restr{D}{(\mathrm{k}, j, m_z, \lambda)} = \mathrm{log}_{10}\left[ \restr{\mathrm{C}_{+\hat{\mathrm{x}}}}{( \mathrm{k}, j, m_z, \lambda)} / \restr{\mathrm{C}_{-\hat{\mathrm{x}}}}{(\mathrm{k}, j, m_z, \lambda)}  \right].\label{eqndirectA}
\end{eqnarray}

We will use Eqs.~(\ref{eqnhelsym2a},\ref{eq:def_coupl},\ref{eqndirectA}), as well as the invariance of $\mathrm{S}$ under $\mathrm{M}_z$ 
\begin{equation}
	\label{eq:invMz}
\mathrm{M}_z^\dagger \mathrm{S} \mathrm{M}_z = \mathrm{S},
\end{equation}

and the transformation of $\sigmapm$ under $\Mz$
\begin{equation}
	\label{eq:mztetm}
	\Mz\sigmapm=q\sigmapm,
\end{equation}
where $q$ is either +1 or -1. Equation~(\ref{eq:mztetm}) follows from the invariance of $\mathrm{S}$ under $\mathrm{M}_z$, whereby the guided modes in the $\pm \hat{\mathrm{x}}$ direction must transform as $\mathrm{M}_z \ket{\sigma_{\pm \hat{\mathrm{x}}}} = e^{\mathrm{i}\varphi_{\pm}} \ket{\sigma_{\pm \hat{\mathrm{x}}}} $. Then, since $\Mz^2=I$, it must be that $e^{\mathrm{i}\varphi_{+}}$ and  $e^{\mathrm{i}\varphi_{-}}$ are equal to either +1 or -1. Finally, since such sign determines the character of the mode upon transformation with $\Mz$, it must be equal for both $\sigmapm$ because they are counter-propagating but otherwise identical modes. We can then write 
\begin{equation}
	\label{eq:e4}
	\begin{split}
		&\restr{\mathrm{C}_{\pm\hat{\mathrm{x}}}}{(\mathrm{k}, j, m_z, \lambda)}  =  \vert \bra{\sigma_{\pm \hat{\mathrm{x}}}} \hspace{2pt} \mathrm{S} \hspace{2pt} \ket{\mathrm{k}\hspace{2pt}j\hspace{2pt}m_z\hspace{2pt}\lambda} \vert^2 \duetoeq{eq:invMz}  \\  
		&\vert \bra{\sigma_{\pm \hat{\mathrm{x}}}} \hspace{2pt} \mathrm{M}_z^\dagger \mathrm{S} \mathrm{M}_z \hspace{2pt} \ket{\mathrm{k}\hspace{2pt}j\hspace{2pt}m_z\hspace{2pt}\lambda} \vert^2\equaldueto{Eqs.~(\ref{eqnhelsym2a},\ref{eq:mztetm})}\\
		&\vert q(-1)^{j+m_z+1} \bra{\sigma_{\pm \hat{\mathrm{x}}}} \hspace{2pt} \mathrm{S} \hspace{2pt} \ket{\mathrm{k}\hspace{2pt}j\hspace{2pt}m_z\hspace{2pt} -\lambda} \vert^2 =  \\
		&\vert \bra{\sigma_{\pm \hat{\mathrm{x}}}} \hspace{2pt} \mathrm{S} \hspace{2pt} \ket{\mathrm{k}\hspace{2pt}j\hspace{2pt}m_z\hspace{2pt} - \lambda} \vert^2 = \restr{\mathrm{C}_{\pm\hat{\mathrm{x}}}}{(\mathrm{k}, j, m_z, - \lambda)}.
	\end{split}
\end{equation}
It then follows that
\begin{equation}
	\label{eq:symcond1}
\restr{D}{(\mathrm{k}, j, m_z, \lambda)} = \restr{D}{(\mathrm{k}, j, m_z, - \lambda)},
\end{equation}
which proves statement 1 above.

We will now use Eqs.~(\ref{eqnhelsym1a},\ref{eq:def_coupl},\ref{eqndirectA}), as well as the invariance of $\mathrm{S}$ under $\mathrm{M}_x$: 
\begin{equation}
\label{eq:invMx}
\mathrm{M}_x^\dagger \mathrm{S} \mathrm{M}_x = \mathrm{S}. 
\end{equation}
	Due to the fact that the power in the waveguide travels from one end to the other, the guided modes are not eigenstates of $\mathrm{M}_x$. Instead, they are transformed into each other as $\mathrm{M}_x \ket{\sigma_{\pm \hat{\mathrm{x}}}} = p \ket{\sigma_{\mp \hat{\mathrm{x}}}} $, with $p$ equal to either +1 or -1. Therefore, we have that:

\begin{equation}
	\begin{split}
		&\restr{\mathrm{C}_{\pm\hat{\mathrm{x}}}}{(\mathrm{k}, j, m_z, \lambda)}  =  \vert \bra{\sigma_{\pm \hat{\mathrm{x}}}} \hspace{2pt} \mathrm{S} \hspace{2pt} \ket{\mathrm{k}\hspace{2pt}j\hspace{2pt}m_z\hspace{2pt}\lambda} \vert^2 \duetoeq{eq:invMx} \\  
		&\scalebox{1}{$ \vert \bra{\sigma_{\pm \hat{\mathrm{x}}}} \hspace{2pt} \mathrm{M}_x^\dagger \mathrm{S} \mathrm{M}_x \hspace{2pt} \ket{\mathrm{k}\hspace{2pt}j\hspace{2pt}m_z\hspace{2pt}\lambda} \vert^2 \duetoeq{eqnhelsym1a}$}\\ 
		& \scalebox{1}{$\vert p(-1)^{(m_z+1)} \bra{\sigma_{\mp \hat{\mathrm{x}}}} \hspace{2pt} \mathrm{S} \hspace{2pt} \ket{\mathrm{k}\hspace{2pt}j\hspace{2pt} -m_z\hspace{2pt} -\lambda} \vert^2 =$} \\
		&\scalebox{1}{$ \vert \bra{\sigma_{\mp \hat{\mathrm{x}}}} \hspace{2pt} \mathrm{S} \hspace{2pt} \ket{\mathrm{k}\hspace{2pt}j\hspace{2pt} -m_z\hspace{2pt} - \lambda} \vert^2 = \restr{\mathrm{C}_{\mp \hat{\mathrm{x}}}}{(\mathrm{k}, j, -m_z, - \lambda)}, $}
\end{split}
\end{equation}

\noindent which implies:

\begin{equation}
	\label{eq:symcond2}
    \restr{D}{(\mathrm{k}, j, m_z, \lambda)} = -\restr{D}{(\mathrm{k}, j, -m_z, -\lambda)}.
\end{equation}

We now combine Eq.~(\ref{eq:symcond1}) and Eq.~(\ref{eq:symcond2}) to obtain that, for waveguides that are invariant under both $\mathrm{M}_x$ and $\mathrm{M}_z$, and when the emitter is located in the xOy plane:
\begin{equation}
	\label{eq:symcond3}
\restr{D}{(\mathrm{k}, j, m_z, \lambda)}= -\restr{D}{(\mathrm{k}, j, -m_z,  \lambda)},
\end{equation}
which proves statement 2 above. Statement 3 is readily shown by particularizing Eq.~(\ref{eq:symcond3}) for $m_z=0$:
\begin{equation}
\restr{D}{(\mathrm{k}, j, 0, \lambda)}=-\restr{D}{(\mathrm{k}, j, 0, \lambda)}\implies \restr{D}{(\mathrm{k}, j, 0, \lambda)}=0.
\end{equation}

\subsection{Electric and magnetic multipoles\label{app:elmagdir}}
We now consider the electric ($\tau=1$) and magnetic ($\tau=-1$) multipoles $\mptau$. They can be written as linear combinations of the helical multipoles by inverting \Eq{eqnheldef1}:
\begin{equation}
	\label{eq:taump}
	\mptau=\frac{\mplambdaplus-\tau\mplambdaminus}{\sqrt{2}}.
\end{equation}

We will now show that, when the electric(magnetic) multipoles couple to the waveguide, their directionality is identical to the one for the helical multipoles with the same $(\mathrm{k},j,m_z)$ numbers.

\begin{equation}
	\label{eq:ss}
	\begin{split}
		&\restr{\mathrm{C}_{\pm\hat{\mathrm{x}}}}{(\mathrm{k}, j, m_z, \tau)}=\\
		&\frac{1}{2}\vert  \sigmapmbra \mathrm{S}  \mplambdaplus-\tau\sigmapmbra \mathrm{S} \mplambdaminus\vert^2=\\ 
		&\frac{1}{2}\left(\vert  \sigmapmbra \mathrm{S}  \mplambdaplus\vert^2+\vert  \sigmapmbra \mathrm{S}\mplambdaminus\vert^2\right)-\\
		&\tau\text{Re}\left\{{\sigmapmbra \mathrm{S}  \mplambdaplus}^*{\sigmapmbra \mathrm{S}  \mplambdaminus}\right\}\duetoeq{eq:e4}\\
		&\vert  \sigmapmbra \mathrm{S}  \mplambdaplus\vert^2-\\
		&\tau\text{Re}\left\{{\sigmapmbra \mathrm{S}  \mplambdaplus}^*{\sigmapmbra \mathrm{S}  \mplambdaminus}\right\},
	\end{split}
\end{equation}
Let us now manipulate the last term in \Eq{eq:ss} 
\begin{equation}
	\begin{split}	
		&	\sigmapmbra \mathrm{S}  \mplambdaminus\duetoeq{eq:invMz}\sigmapmbra \Mz^\dagger\mathrm{S}\Mz \mplambdaminus\\
		& \equaldueto{\Eq{eqnhelsym2a},\Eq{eq:mztetm}}q(-1)^{j+m_z+1}\sigmapmbra \mathrm{S}  \mplambdaplus,\\
	\end{split}
\end{equation}
and substitute it back into \Eq{eq:ss}:
\begin{equation}
	\label{eq:final}
\begin{split}
		&\restr{\mathrm{C}_{\pm\hat{\mathrm{x}}}}{(\mathrm{k}, j, m_z, \tau)}=\vert  \sigmapmbra \mathrm{S}  \mplambdaplus\vert^2-\\
		&\tau\text{Re}\left\{q(-1)^{j+m_z+1}|{\sigmapmbra \mathrm{S}  \mplambdaplus}|^2\right\}=\\
		&(1+\tau q(-1)^{j+m_z})\vert  \sigmapmbra \mathrm{S}  \mplambdaplus\vert^2.
\end{split}
\end{equation}
Two important conclusions can be reached from \Eq{eq:final}. First, when $\tau q(-1)^{j+m_z}=-1$ the emission {\em cannot} couple into the waveguide at all. The reciprocal version of this selection rule can be found in \cite[Tab. I]{transversecouplingevanescentwaves} for the decomposition of a single evanescent plane-wave into multipoles with well-defined transverse angular momentum. And second, when the selection rule allows the coupling, the directionality is identical to the one for the helical multipoles:

\begin{equation}
	\label{eq:finalfinal}
	\begin{split}
		&\restr{D}{(\mathrm{k}, j, m_z, \tau)} = \mathrm{log}_{10}\left[ \restr{\mathrm{C}_{+\hat{\mathrm{x}}}}{( \mathrm{k}, j, m_z, \tau)} / \restr{\mathrm{C}_{-\hat{\mathrm{x}}}}{(\mathrm{k}, j, m_z, \tau)}  \right]\\
		&\duetoeq{eq:final} \mathrm{log}_{10}\left[\frac{2 \vert  \sigmaplusbra \mathrm{S}  \mplambda\vert^2}{2 \vert  \sigmaminusbra\mathrm{S}  \mplambda\vert^2}\right]\duetoeq{eq:ddef}\restr{D}{(\mathrm{k}, j, m_z, \lambda)}.
	\end{split}
\end{equation}

The same directionality is also featured by general combinations of helical multipoles obtained with an arbitrary complex value of $\tau$ in \Eq{eq:taump}. The only difference in \Eq{eq:final} is a substitution $\tau\rightarrow\text{Re}\{\tau\}$, which does not affect \Eq{eq:finalfinal}.

\subsection{A different angular momentum axis\label{app:jy}}
We consider an emission expressed as the sum of helical multipoles with well-defined angular momentum along the $\yhat$ axis:
\begin{equation}
	\psiket = \sum_{\mathrm{k},j,m_y,\lambda}\alpha(\mathrm{k},j,m_y,\lambda)\mplambdamy,
\end{equation}
where the sum in $\mathrm{k}$ is meant to represent an integral. The emitter is placed on the XY plane. We assume that the coefficients $\alpha(\mathrm{k},j,m_y,\lambda)$ are equal to zero for odd(even) $m_y$. Under this assumption, the emission is an eigenstate of the combined $\Mz\Mx$ transformation:
\begin{equation}
	\label{eq:psimy}
	\Mz\Mx\psiket = s\psiket,
\end{equation}
where $s=1(-1)$ when all coefficients $\alpha(\mathrm{k},j,m_y,\lambda)$ are equal to zero for odd(even) $m_y$. Equation~(\ref{eq:psimy}) follows from considering Eqs.~(\ref{eq:b6},\ref{eq:b7}) after a rotation by $-\pi/2$ along the $\xhat$ axis which maps $y$ to $-z$ and $z$ to $y$. Together, the two equations imply that $\Mz\Mx\mplambdamy=(-1)^{m_y}\mplambdamy$.

We now consider the power coupled towards each direction
\begin{equation}
\restr{\mathrm{C}_{\pm\hat{\mathrm{x}}}}{\psiket} = \vert \bra{\sigma_{\pm \hat{\mathrm{x}}}} \hspace{2pt} \mathrm{S} \hspace{2pt} \psiket \vert^2
\end{equation}
and apply the consequences of the combined transformation
\begin{equation}
	\label{eq:jarl}
	\begin{split}
		&\restr{\mathrm{C}_{\pm\hat{\mathrm{x}}}}{\psiket} = \\
		&\vert \bra{\sigma_{\pm \hat{\mathrm{x}}}} \hspace{2pt} \left(\Mz\Mx\right)^\dagger\mathrm{S}\Mz\Mx \hspace{2pt} \psiket \vert^2=\vert pqs\bra{\sigma_{\mp \hat{\mathrm{x}}}} \hspace{2pt} \mathrm{S}\hspace{2pt} \psiket \vert^2
	\end{split}
\end{equation}
to show that the directionality is identically zero since $pqs$ is equal to either +1 or -1 . The second equality in \Eq{eq:jarl} follows from \Eq{eq:psimy}, and the arguments around Eqs.~(\ref{eq:mztetm},\ref{eq:invMx}).

\section{Plane-wave spectrum of multipoles of well-defined helicity and the directionality of its evanescent part\label{app:PWS}}

In this Section, we will examine the plane-wave expansion of the emission of a multipolar source with well-defined helicity. It is our purpose to investigate the origin of the exponential dependence of the directionality on the transverse component of the angular momentum. We will show that the exponential dependence has its cause in an intrinsic characteristic of the evanescent part of the angular spectrum of the emission: The ratio of the energy flux densities carried by evanescent plane-waves with opposite $\mathrm{k}_x$ is proportional to a term that has an exponential dependence on the transverse angular momentum $m_z$. It is that $m_z$-driven asymmetry in the energy flux that translates to the directional coupling.

\subsection{Plane-wave spectrum representation of a multipolar emission for the half-space that is transverse to its quantization axis}

In Ref.~\onlinecite{multDevWolf} [Eqs.~(B1a, B1b)], Devaney and  Wolf expand the fields of multipoles of well-defined parity in a series of plane-waves containing both propagating and evanescent components. By using our introduced conventions, normalizations, and from the definition of the multipoles of well-defined helicity [Eq.~(\ref{eqnheldef1})], we can reach the following formula that expands the helical multipoles $\ket{\mathrm{k}\hspace{2pt}j\hspace{2pt}m_z\hspace{2pt}\lambda}$ as an integral series of plane-waves that is valid for the z>0 half-space:

\begin{eqnarray}
    \ket{\mathrm{k}\hspace{2pt}j\hspace{2pt}m_z\hspace{2pt}\lambda}&\equiv&\frac{1}{2\pi\mathrm{i}^{j-1}}\int_{C^{+\hat{z}}_{\phi_{\hat{\mathrm{k}}}}}\mathrm{d}\phi_{\hat{\mathrm{k}}} \int_{C^{+\hat{z}}_{\theta_{\hat{\mathrm{k}}}}}\mathrm{sin}\theta_{\hat{\mathrm{k}}}\mathrm{d}\theta_{\hat{\mathrm{k}}}\mathrm{A}_{\lambda,m_z j}(\hat{\mathrm{k}}) \hat{\mathrm{e}}_{\lambda}(\hat{\mathrm{k}})e^{\mathrm{i}\mathbf{k}\cdot\mathbf{r}},\nonumber\\
    &\equiv&\frac{1}{2\pi\mathrm{i}^{j-1}}\iint\limits_{-\infty}^{+\infty}\frac{\mathrm{dk}_x\mathrm{dk}_y}{\mathrm{k}\sqrt{\mathrm{k}^2-\mathrm{k}_x^2-\mathrm{k}_y^2}}\mathrm{A}_{\lambda,m_z j}(\hat{\mathrm{k}}) \hat{\mathrm{e}}_{\lambda}(\hat{\mathrm{k}})e^{\mathrm{i}\mathbf{k}\cdot\mathbf{r}},\nonumber\\
    &&\hspace{133pt}\mathrm{for}\hspace{6pt} \mathrm{z> 0},\label{eqnPWSdecomp1}
\end{eqnarray}

\noindent The positive half space (z>0) is defined relative to the position of the emitter (see Fig. 1 of the main manuscript). The normal vector of the interface defining the half-space points to the direction of the quantization axis of the emitter. The wavevector direction of each plane-wave component in the definition above is given by: 

\begin{eqnarray}
    \hat{\mathrm{k}}(\theta_{\hat{\mathrm{k}}},\phi_{\hat{\mathrm{k}}})&=&(\mathrm{k}_x\hat{\mathrm{x}}+\mathrm{k}_y\hat{\mathrm{y}}+\mathrm{k}_z\hat{\mathrm{z}})/\mathrm{k}\nonumber\\
    &=&\hat{\mathrm{x}}\mathrm{sin}\theta_{\hat{\mathrm{k}}}\mathrm{cos}\phi_{\hat{\mathrm{k}}}+\hat{\mathrm{y}}\mathrm{sin}\theta_{\hat{\mathrm{k}}}\mathrm{sin}\phi_{\hat{\mathrm{k}}}+\hat{\mathrm{z}}\mathrm{cos}\theta_{\hat{\mathrm{k}}}.\label{direct_kspace}
\end{eqnarray}

\noindent The polar and the azimuthal angles of propagation are defined by: 

\begin{eqnarray}
	\theta_{\hat{\mathrm{k}}}&=&\mathrm{arccos}(\mathrm{k}_z/\mathrm{k})=-\mathrm{i}\hspace{2pt}\mathrm{ln}\left[\mathrm{k}_z/\mathrm{k}+\mathrm{i}\sqrt{1-(\mathrm{k}_z/\mathrm{k})^2}\right],\label{eqntheta}\\
    \phi_{\hat{\mathrm{k}}}&=&\mathrm{arctan}(\mathrm{k}_x,\mathrm{k}_y)=-\mathrm{i}\hspace{2pt}\mathrm{ln}\left[\frac{\mathrm{k}_x+\mathrm{ik}_y}{\sqrt{\mathrm{k}_x^2+\mathrm{k}_y^2}}\right],\label{eqnphi}
\end{eqnarray}

\noindent and their integration contour at the integral above is $C^{+\hat{z}}_{\theta_{\hat{\mathrm{k}}}}=[0,\frac{\pi}{2}-\mathrm{i}\infty]$ and $C^{+\hat{z}}_{\phi_{\hat{\mathrm{k}}}}=[0,2\pi]$  respectively. The complex polar angles $\theta_{\hat{\mathrm{k}}}$ account for the evanescent part of the plane-wave spectrum. The latter formulas give the analytic continuation of the polar and azimuthal angles in the complex plane as a function of the Cartesian components of the wavevector $\mathbf{k}$.  $\mathrm{k}_z(\mathrm{k},\mathrm{k}_x,\mathrm{k}_y)=\sqrt{\mathrm{k}^2-\mathrm{k}_x^2-\mathrm{k}_y^2}$ is a restricted variable that takes values on the positive real(imaginary) part of the z-axis for propagating(evanescent) plane-waves that propagate(decay) along the $+\hat{\mathrm{z}}$ direction.

The spectral amplitudes $\mathrm{A}_{\lambda,m_z j}(\hat{\mathrm{k}})$ are given by Eq.~(\ref{eqnspectrum}) and Eq.~(\ref{eqnpolvector}) gives the polarization vector:

\begin{eqnarray}
    \hat{\mathrm{e}}_{\lambda}(\hat{\mathrm{k}})&=&\frac{\hat{\mathrm{x}}}{\sqrt{2}}\left(-\lambda\mathrm{cos}\theta_{\hat{\mathrm{k}}} \mathrm{cos}\phi_{\hat{\mathrm{k}}} +\mathrm{i}\hspace{2pt}\mathrm{sin}\phi_{\hat{\mathrm{k}}}\right)\nonumber\\
    &+&\frac{\hat{\mathrm{y}}}{\sqrt{2}}\left(-\lambda\mathrm{cos}\theta_{\hat{\mathrm{k}}} \mathrm{sin}\phi_{\hat{\mathrm{k}}}-\mathrm{i}\hspace{2pt}\mathrm{cos}\phi_{\hat{\mathrm{k}}}\right)\nonumber\\
    &+&\frac{\hat{\mathrm{z}}}{\sqrt{2}}\lambda\mathrm{sin}\theta_{\hat{\mathrm{k}}}.\label{eqnpolvector2}
\end{eqnarray}

\noindent It is important to note that each plane wave $\hat{\mathrm{e}}_{\lambda}(\hat{\mathrm{k}})e^{\mathrm{i}\mathbf{k}\cdot\mathbf{r}}$, independent of whether it is a propagating or an evanescent plane wave, is divergent-free: $\hat{\mathrm{e}}_{\lambda}(\hat{\mathrm{k}})\cdot\hat{\mathrm{k}}=0$. Each plane wave also constitutes an eigenstate of the helicity operator with eigenvalue $\lambda$. The plane-wave spectrum of a multipole with $\lambda=+1(-1)$ is purely composed out of left-handed(right-handed) circularly polarized plane waves. \textit{The helicity $\lambda$ defines the handedness of the polarization in momentum space}.

However, we also note that, for the evanescent part of the spectrum, apart from the norm of the unit wavevectors $\hat{\mathrm{k}}$ [see Eq.~(\ref{direct_kspace})], also the norm of the corresponding polarization vector $\hat{\mathrm{e}}_{\lambda}(\hat{\mathrm{k}})$ stops being unitary. From Eq.~(\ref{eqnpolvector2}) we have that: 

\begin{eqnarray}
	\left|\hat{\mathrm{e}}_{\lambda}(\hat{\mathrm{k}})\right|&=&\mathrm{cosh}\left( \mathrm{Im}\left\{ \theta_{\hat{\mathrm{k}}} \right\} \right)\mathrm{cosh}\left(\mathrm{Im}\left\{\phi_{\hat{\mathrm{k}}}\right\} \right)\nonumber\\
	&+&\lambda\mathrm{cos}\left(\mathrm{Re}\left\{\theta_{\hat{\mathrm{k}}}\right\} \right)\mathrm{sinh}\left(\mathrm{Im}\left\{\phi_{\hat{\mathrm{k}}}\right\} \right).\label{eqnabspol}
\end{eqnarray}

\noindent  For complex angles, the polarization vectors of opposite helicity stop being orthogonal in the usual sense: $\hat{\mathrm{e}}_{\lambda}(\hat{\mathrm{k}})\cdot\hat{\mathrm{e}}^*_{\lambda'}(\hat{\mathrm{k}})\neq\delta_{\lambda\lambda'}$. Instead, we have the following orthogonality property that is also valid for complex angles of propagation: $\hat{\mathrm{e}}_{\lambda}(\hat{\mathrm{k}})\cdot\hat{\mathrm{e}}_{-\lambda'}(\hat{\mathrm{k}})=-\delta_{\lambda\lambda'}$. 

Before we move further on, let us introduce a couple of other properties of the polarization vector that will be useful later:

\begin{eqnarray}
	|\hat{\mathrm{e}}_{\lambda}(\mathrm{k}_x,\mathrm{k}_z)|&=& |\hat{\mathrm{e}}_{-\lambda}(-\mathrm{k}_x,\mathrm{k}_z)|, \label{eqnAbsPolSym1}\\
	|\hat{\mathrm{e}}_{\lambda}(\mathrm{k}_x,\mathrm{k}_z)|&=& |\hat{\mathrm{e}}_{-\lambda}(\mathrm{k}_x,-\mathrm{k}_z)|.
	\label{eqnAbsPolSym2}
\end{eqnarray}

\noindent Equation~(\ref{eqnAbsPolSym1}) follows because $\theta_{\hat{\mathrm{k}}}$ does not depend on $\mathrm{k}_x$, and $\mathrm{Im}\left\{\phi_{\hat{\mathrm{k}}}(\mathrm{k}_x,\mathrm{k}_z)\right\}=-\mathrm{Im}\left\{\phi_{\hat{\mathrm{k}}}(-\mathrm{k}_x,\mathrm{k}_z)\right\}=\mathrm{ln}\left|\sqrt{\mathrm{k}_x^2+\mathrm{k}_y^2}\right|-\mathrm{ln}\left|\mathrm{k}_x+\mathrm{i}\hspace{2pt}\mathrm{k}_y\right|$. Equation~(\ref{eqnAbsPolSym2}) follows because $\theta_{\hat{\mathrm{k}}}(\mathrm{k}_x,\mathrm{k}_z)=\pi-\theta_{\hat{\mathrm{k}}}(\mathrm{k}_x,-\mathrm{k}_z)$ and $\phi_{\hat{\mathrm{k}}}(\mathrm{k}_x,\mathrm{k}_z)=\phi_{\hat{\mathrm{k}}}(\mathrm{k}_x,-\mathrm{k}_z)$.

So, Eq.~(\ref{eqnPWSdecomp1}) gives the plane-wave expansion that describes the fields in the z>0 half-space. However, in our case we are interested in the plane-wave expansion for the y<0 half-space, because this is the half-space that hosts the waveguide (see Fig.~\ref{fig:fig1} of the main text). To take the plane-wave expansion that describes the radiated fields in an arbitrary half-space, we proceed as follows:

We begin by expressing the multipolar emission $\ket{\mathrm{k}\hspace{2pt}j\hspace{2pt}m_{z}\hspace{2pt}\lambda}$ as a superposition of multipoles $\ket{\mathrm{k}\hspace{2pt}j\hspace{2pt}m_{z'}\hspace{2pt}\lambda}$ with well-defined angular momentum along the z-axis, z', of a rotated coordinate frame that is given by a z-y-z rotation of the original one under the Euler angles $(\alpha,\beta,\gamma)$. This inverse rotation of the multipoles is done by making use of the Wigner D-Matrix \cite{tung1985group}. We formulate this here for arbitrary angles of $\alpha$, $\beta$, and $\gamma$, but afterwards, of course, specific values are considered to account for the specific rotation of the coordinate system we are interested in. As a second step, we apply Eq.~(\ref{eqnPWSdecomp1}) to get the plane-wave expansion for the z'>0 half-space -which shall be the half-space that hosts the waveguide (y<0 in our case)-:

\begin{eqnarray}
    \ket{\mathrm{k}\hspace{2pt}j\hspace{2pt}m_z\hspace{2pt}\lambda}
    &=&\sum_{m_{z'}=-j}^j\mathrm{D}^{j}_{m_{z'}m_{z}}(-\gamma,-\beta,-\alpha)\ket{\mathrm{k}\hspace{2pt}j\hspace{2pt}m_{z'}\hspace{2pt}\lambda}\nonumber\\
    &\equiv&\sum_{m_{z'}=-j}^j\mathrm{D}^{j}_{m_{z'}m_{z}}(-\gamma,-\beta,-\alpha)\times\nonumber\\
    &\times&\frac{1}{2\pi\mathrm{i}^{j-1}}\iint\limits_{-\infty}^{+\infty}\frac{\mathrm{dk}_{x'}\mathrm{dk}_{y'}}{\mathrm{k}\sqrt{\mathrm{k}^2-\mathrm{k}_{x'}^2-\mathrm{k}_{y'}^2}}\mathrm{A}_{\lambda,m_{z'} j}(\hat{\mathrm{k}}') \hat{\mathrm{e}}_{\lambda}(\hat{\mathrm{k}}')e^{\mathrm{i}\mathbf{k}'\cdot\mathbf{r}'},\nonumber\\
    &&\hspace{114pt}\mathrm{for}\hspace{6pt} \mathrm{z'}> 0.\label{eqnwignerd1}
    \end{eqnarray}
    
    \noindent Next, we proceed with the following change of variables: $\mathbf{r}'=\mathrm{R}\mathbf{r}$, $\mathbf{k}'=\mathrm{R}\mathbf{k}$, where $\mathrm{R}(\alpha,\beta,\gamma)=[\mathrm{R}_{{x'}} \hspace{3pt}  \mathrm{R}_{{y'}} \hspace{3pt} \mathrm{R}_{{z'}} ]^\mathrm{T}$ is a 3x3 matrix that rotates the original coordinate system under the Euler angles $(\alpha,\beta,\gamma)$ corresponding to a z-y-z rotation, so that the new z-axis, $\hat{\mathrm{z}}'$, is along the direction that defines the interior of the half-space of our interest. R is a real unitary matrix having the property $\mathrm{R^{-1}(\alpha,\beta,\gamma)=R^T(\alpha,\beta,\gamma)=R(-\gamma,-\beta,-\alpha)}$, which means that:\ $\mathbf{k}'\cdot\mathbf{r}'=[\mathbf{k}^\mathrm{T}\mathrm{R^T}][\mathrm{R}\mathbf{r}]=\mathbf{k}^\mathrm{T}\mathbf{r}=\mathbf{k}\cdot\mathbf{r}$. Applying the above and  rearranging the sums gives:

    \begin{eqnarray}
    \ket{\mathrm{k}\hspace{2pt}j\hspace{2pt}m_z\hspace{2pt}\lambda}
    &=&\frac{1}{2\pi\mathrm{i}^{j-1}}\iint\limits_{-\infty}^{+\infty}\frac{\mathrm{d}[\mathrm{R}_{x'}\mathbf{k}]\mathrm{d}[\mathrm{R}_{y'}\mathbf{k}]}{\mathrm{k}\sqrt{\mathrm{k}^2-[\mathrm{R}_{x'}\mathbf{k}]^2-[\mathrm{R}_{y'}\mathbf{k}]^2}}e^{\mathrm{i}\mathbf{k}\cdot\mathbf{r}}\times\nonumber\\
    &\times&\left[\sum_{m_{z'}=-j}^j\mathrm{D}^{j}_{m_{z'}m_{z}}(-\gamma,-\beta,-\alpha)\mathrm{A}_{\lambda,m_{z'} j}(\hat{\mathrm{k}}') \hat{\mathrm{e}}_{\lambda}(\hat{\mathrm{k}}')\right],\nonumber\\
    &&\hspace{114pt}\mathrm{for}\hspace{6pt} \mathrm{R}_{z'}\mathbf{r}> 0. \label{eqnD9}
    \end{eqnarray}
    
     \noindent As a last but crucial step we need to calculate the sum inside the square brackets of the above formula. For this, one needs to notice -by looking at the definitions of Eqs.~(\ref{eqnspectrum},\ref{eqnpolvector}) and the representations of the nabla operators in a spherical coordinate system- that:

     \begin{eqnarray}
     \mathrm{A}_{\lambda,m_{z} j}(\hat{\mathrm{k}}) \hat{\mathrm{e}}_{\lambda}(\hat{\mathrm{k}}) &=& \hat{\mathbf{O}}\hspace{2pt}\mathrm{Y}_j^{m_z}(\hat{\mathrm{k}}),\nonumber\\
      \end{eqnarray}
      
    \noindent where $\mathrm{Y}_j^{m_z}(\hat{\mathrm{k}})=\gamma_j^{m_z}\mathrm{P}_j^{m_z}(\mathrm{cos}\theta_{\hat{\mathrm{k}}})e^{\mathrm{i}m_z\phi_{\hat{\mathrm{k}}}}$ are the spherical harmonics and the operator $\hat{\mathbf{O}}$ is defined as:
    
    \begin{eqnarray}
    \hat{\mathbf{O}}&=&\frac{-\mathrm{i}\nabla_{\mathrm{k}}\times\left[\mathbf{k}(\cdot)\right]+\lambda\mathrm{k}\nabla_{\mathrm{k}}(\cdot)}{\sqrt{2}},
    \end{eqnarray}
    
    \noindent with the subscript k at the nablas implying operation in the k-space. Applying the same inverse rotation to the scalar spherical harmonics using the Wigner D-matrices, as we did in Eq.~(\ref{eqnwignerd1}) for the multipoles, gives:
    
    \begin{eqnarray}
    \mathrm{A}_{\lambda,m_{z} j}(\hat{\mathrm{k}}) \hat{\mathrm{e}}_{\lambda}(\hat{\mathrm{k}}) &=& \hat{\mathbf{O}}\hspace{2pt}\mathrm{Y}_j^{m_z}(\hat{\mathrm{k}})\nonumber\\
    &=&\hat{\mathbf{O}} \hspace{2pt}\left[\sum\limits_{m_{z'}=-j}^j\mathrm{D}^{j}_{m_{z'}m_{z}}(-\gamma,-\beta,-\alpha) \mathrm{Y}_j^{m_{z'}}(\hat{\mathrm{k}}')\right]\nonumber\\
    &=&\sum\limits_{m_{z'}=-j}^j\mathrm{D}^{j}_{m_{z'}m_{z}}(-\gamma,-\beta,-\alpha)\hat{\mathbf{O}} \hspace{2pt}\mathrm{Y}_j^{m_{z'}}(\hat{\mathrm{k}}')\nonumber\\
    &=&\sum\limits_{m_{z'}=-j}^j\mathrm{D}^{j}_{m_{z'}m_{z}}(-\gamma,-\beta,-\alpha) \mathrm{A}_{\lambda,m_{z'} j}(\hat{\mathrm{k}}') \hat{\mathrm{e}}_{\lambda}(\hat{\mathrm{k}}').\nonumber\\
    \end{eqnarray}
     
    \noindent Substituting this expression into Eq.~(\ref{eqnD9}) finally gives us the formula for the momentum space representation of the radiation of a helical multipole valid for an arbitrary half-space $\mathrm{R}_{z'}\mathbf{r}> 0$: 
     
    \begin{eqnarray}
    \ket{\mathrm{k}\hspace{2pt}j\hspace{2pt}m_z\hspace{2pt}\lambda}
    &=&\frac{1}{2\pi\mathrm{i}^{j-1}}\iint\limits_{-\infty}^{+\infty}\frac{\mathrm{d}[\mathrm{R}_{x'}\mathbf{k}]\mathrm{d}[\mathrm{R}_{y'}\mathbf{k}]\mathrm{A}_{\lambda,m_{z} j}(\hat{\mathrm{k}}) \hat{\mathrm{e}}_{\lambda}(\hat{\mathrm{k}})e^{\mathrm{i}\mathbf{k}\cdot\mathbf{r}}}{\mathrm{k}\sqrt{\mathrm{k}^2-[\mathrm{R}_{x'}\mathbf{k}]^2-[\mathrm{R}_{y'}\mathbf{k}]^2}},\nonumber\\
    &&\hspace{114pt}\mathrm{for}\hspace{6pt} \mathrm{R}_{z'}\mathbf{r}> 0\label{eqnrotangspec}
\end{eqnarray}

In our specific case, a rotation matrix R that transforms -$\hat{\mathrm{y}}$ into $\hat{\mathrm{z}}'$ can be the following:

\begin{eqnarray}
   \mathrm{R}(\alpha,\beta,\gamma)=\mathrm{R}(3\pi/2,\pi/2,\pi/2)=\begin{bmatrix}\mathrm{R}_{{x'}} \\ \mathrm{R}_{{y'}} \\ \mathrm{R}_{{z'}} \end{bmatrix}=
   \begin{bmatrix}1 & 0 & 0\\
                  0 & 0 & 1\\
                  0 &-1 & 0\end{bmatrix}. \nonumber\\\label{eqnrotmat}
\end{eqnarray}

\noindent Finally, by substituting the above into it, Eq.~(\ref{eqnrotangspec})  takes the following specific form:

\begin{eqnarray}
    \ket{\mathrm{k}\hspace{2pt}j\hspace{2pt}m_z\hspace{2pt}\lambda}&=&\frac{1}{2\pi\mathrm{i}^{j-1}}\iint\limits_{-\infty}^{+\infty}\frac{\mathrm{dk}_{x}\mathrm{dk}_{z}}{-\mathrm{kk}_{y}}\mathrm{A}_{\lambda,m_{z} j}(\hat{\mathrm{k}}) \hat{\mathrm{e}}_{\lambda}(\hat{\mathrm{k}})e^{\mathrm{i}\mathbf{k}\cdot\mathbf{r}},\nonumber\\
    &&\hspace{105pt}\mathrm{for}\hspace{6pt} \mathrm{y}< 0,\label{eqnTPWS}
\end{eqnarray}

\noindent where $\mathrm{k}_y(\mathrm{k},\mathrm{k}_x,\mathrm{k}_z)=-\sqrt{\mathrm{k}^2-\mathrm{k}_x^2-\mathrm{k}_z^2}$ is now the restricted variable that takes values on the negative real(imaginary) part of the y-axis for propagating(evanescent) plane-waves that propagate(decay) along the $-\hat{\mathrm{y}}$ direction. $\mathrm{A}_{\lambda,m_{z} j}(\hat{\mathrm{k}})$ and $\hat{\mathrm{e}}_{\lambda}(\hat{\mathrm{k}})$ are analytic functions of $\theta_{\hat{\mathrm{k}}}$ and $\phi_{\hat{\mathrm{k}}}$, and, therefore, we can use Eqs.~(\ref{eqntheta},\ref{eqnphi}) and have access to their analytic continuation in the complex plane. We see from Eq.~(\ref{eqnrotangspec}) that the angular spectrum function $\mathrm{A}_{\lambda,m_{z} j}(\hat{\mathrm{k}})\hat{\mathrm{e}}_{\lambda}(\hat{\mathrm{k}})$ determines the plane-wave expansion of the multipole for an arbitrary half-space. One only needs to modify appropriately the integration contour of the polar and azimuthal angles of propagation in the complex plane to account for the relevant propagating and evanescent part of the spectrum.

 So, Eq.~(\ref{eqnTPWS}) accounts for the transverse plane-wave spectrum of the multipolar emission. That is the plane-wave expansion valid in the half-space y<0 that is transverse to the quantization axis of the emitter (the z-axis) and hosts the waveguide. Next, we will make use of this formula to study the directionality of the evanescent part of the transverse plane-wave spectrum of such helical multipolar emissions.

\subsection{Directionality of the evanescent part of the transverse plane-wave spectrum of the multipolar emission}
Let us now consider the coupling of the emission of a specific multipole $\ket{\mathrm{k}\hspace{2pt}j\hspace{2pt}m_z\hspace{2pt}\lambda}$ into the waveguide on the base of its transverse plane-wave decomposition that we just calculated. We are going to show that the ratio of the energy flux densities carried by the evanescent plane-waves of the transverse multipolar spectrum with opposite $\mathrm{k}_x$ is proportional to a term that has an exponential dependence on the transverse angular momentum $m_z$. Then we will argue that this $m_z$-driven asymmetry in the energy flux density is the main origin of the directionality of the coupling.

We start by showing that only the evanescent plane-waves in the decomposition of the emission can couple power into the guided mode of the waveguide. This follows from the translation-invariance of the waveguide along $\hat{\mathrm{x}}$, which imposes the conservation of the $x$ component of momentum, and makes it impossible for any plane-wave with $\mathrm{k}_x\neq\pm\beta$ to couple into the modes. Then, since $\beta$, the propagation constant of the mode, is larger than the wavenumber outside the waveguide, $\beta=|\mathrm{k}_x|>\mathrm{k}$, it follows that all the contributing plane-waves will be evanescent. Only the plane-wave components of the emission with $\mathrm{k}_x=+\beta$ ($\mathrm{k}_x=-\beta$) -and with varying $\mathrm{k}_z$- can couple power to the guided mode propagating towards the $+\hat{\mathrm{x}}(-\hat{\mathrm{x}})$ direction.
		
Then, with $\mathrm{k}_x$ fixed to either $+\beta$ or $-\beta$, and for fixed $\mathrm{k}_z$ also, a single plane-wave $\mathrm{A}_{\lambda,m_{z} j}(\hat{\mathrm{k}}) \hat{\mathrm{e}}_{\lambda}(\hat{\mathrm{k}})e^{\mathrm{i}\mathbf{k}\cdot\mathbf{r}}$ of the transverse spectrum given by Eq.~(\ref{eqnTPWS}) is chosen for each direction. Evanescent plane-waves do not carry power along the direction of their decay (which is towards the negative y-axis in our case), but they are capable of carrying power along some direction perpendicular to their decay axis. By making use of Eq.~(\ref{eqnpolvector2}) and after some straightforward algebra, we can show that the energy flux density (norm of the real part of the Poynting vector) that such chosen evanescent plane-waves carry is equal to $\left|\mathrm{A}_{\lambda,m_z j}(\hat{\mathrm{k}})\hat{\mathrm{e}}_{\lambda}(\hat{\mathrm{k}})\right|^2/2\mathrm{Z}$. Therefore, the logarithm of the ratio of their energy flux density is given by: 
\begin{equation}
		{R}_{\lambda,m_z j}(\mathrm{k}_z) =	\mathrm{log}_{10}\left[\frac{\left|\mathrm{A}_{\lambda,m_z j}(\mathrm{k}_x=+\beta,\mathrm{k}_z)\hat{\mathrm{e}}_{\lambda}(\mathrm{k}_x=+\beta,\mathrm{k}_z)\right|^2}{\left|\mathrm{A}_{\lambda,m_z j}(\mathrm{k}_x=-\beta,\mathrm{k}_z)\hat{\mathrm{e}}_{\lambda}(\mathrm{k}_x=-\beta,\mathrm{k}_z)\right|^2}\right].\label{eqnratioPW}
	\end{equation}
	
We now use Eqs.~(\ref{eqnspectrum}, \ref{eqntheta}, \ref{eqnphi}, \ref{eqnabspol}, \ref{eqnAbsPolSym1}) to decompose Eq.~(\ref{eqnratioPW}) into two terms:

\begin{equation}
\label{eqnPWdirectA}
	\begin{split}
		&{R}_{\lambda,m_z j}(\mathrm{k}_z) = 2\mathrm{log}_{10}\left[\left|\frac{e^{\mathrm{i}m_z\phi_{\hat{\mathrm{k}}}(\mathrm{k}_x=+\beta,\mathrm{k}_z)}}{e^{\mathrm{i}m_z\phi_{\hat{\mathrm{k}}}(\mathrm{k}_x=-\beta,\mathrm{k}_z)}}\right|\frac{|\hat{\mathrm{e}}_{\lambda}(\mathrm{k}_x=+\beta,\mathrm{k}_z)|}{|\hat{\mathrm{e}}_{\lambda}(\mathrm{k}_x=-\beta,\mathrm{k}_z)|}\right] \\
     &= 2m_z\mathrm{log}_{10}\left[\left|\frac{e^{\mathrm{i}\phi_{\hat{\mathrm{k}}}(\mathrm{k}_x=+\beta,\mathrm{k}_z)}}{e^{\mathrm{i}\phi_{\hat{\mathrm{k}}}(\mathrm{k}_x=-\beta,\mathrm{k}_z)}}\right|\right]
     + 2\lambda\mathrm{log}_{10}\left[\frac{|\hat{\mathrm{e}}_{+}(\mathrm{k}_x=+\beta,\mathrm{k}_z)|}{|\hat{\mathrm{e}}_{+}(\mathrm{k}_x=-\beta,\mathrm{k}_z)|}\right]\\
     &= 2m_z\hspace{1pt}f(\mathrm{k}_z)+2\lambda g(\mathrm{k}_z),
	\end{split}
\end{equation}

\noindent where we have defined:

\begin{eqnarray}
    f(\mathrm{k}_z)&=&\mathrm{log}_{10}\left[\left|\frac{e^{\mathrm{i}\phi_{\hat{\mathrm{k}}}(\mathrm{k}_x=+\beta,\mathrm{k}_z)}}{e^{\mathrm{i}\phi_{\hat{\mathrm{k}}}(\mathrm{k}_x=-\beta,\mathrm{k}_z)}}\right|\right]\\
    &=&\mathrm{log}_{10}\left[\left|\frac{\frac{\beta}{\mathrm{k}}-\mathrm{i}\sqrt{1-\left(\frac{\beta}{\mathrm{k}}\right)^2-\left(\frac{\mathrm{k}_z}{\mathrm{k}}\right)^2}}{\frac{\beta}{\mathrm{k}}+\mathrm{i}\sqrt{1-\left(\frac{\beta}{\mathrm{k}}\right)^2-\left(\frac{\mathrm{k}_z}{\mathrm{k}}\right)^2}}\right|\right],\nonumber
\end{eqnarray}

\noindent and:

\begin{eqnarray}
    g(\mathrm{k}_z)=\mathrm{log}_{10}\left[\frac{|\hat{\mathrm{e}}_{+}(\mathrm{k}_x=+\beta,\mathrm{k}_z)|}{|\hat{\mathrm{e}}_{+}(\mathrm{k}_x=-\beta,\mathrm{k}_z)|}\right].
\end{eqnarray}

 We see that ${R}_{\lambda,m_z j}$ is the sum of two terms: one that is proportional to the transverse angular momentum $m_z$ and another one that is proportional to the helicity $\lambda$. Apart from the fact that both are functions of $\mathrm{k}_z$, there is something to say about the weighting functions of those two terms. On the one hand, we have that $f(\mathrm{k}_z)$, the weighting function of the $2m_z$-dependent term, is always positive since $\beta>\mathrm{k}$ and also has an even symmetry: $f(\mathrm{k}_z)=f(-\mathrm{k}_z)$ .  On the other hand, because of Eqs.~(\ref{eqnAbsPolSym1}, \ref{eqnAbsPolSym2}), $g(\mathrm{k}_z)$, the weighting function of the $2\lambda$-dependent term, is a function with odd symmetry: $g(\mathrm{k}_z)=-g(-\mathrm{k}_z)$. Both functions have singularities at $|\mathrm{k}_z|=\mathrm{k}$ and approach zero in the limit of $|\mathrm{k}_z|\rightarrow\infty$. In Fig.~\ref{fig:fig4} we plot the two functions for the case of $\beta/\mathrm{k}=2.26$.

\begin{figure}
\includegraphics[width=8.5cm]{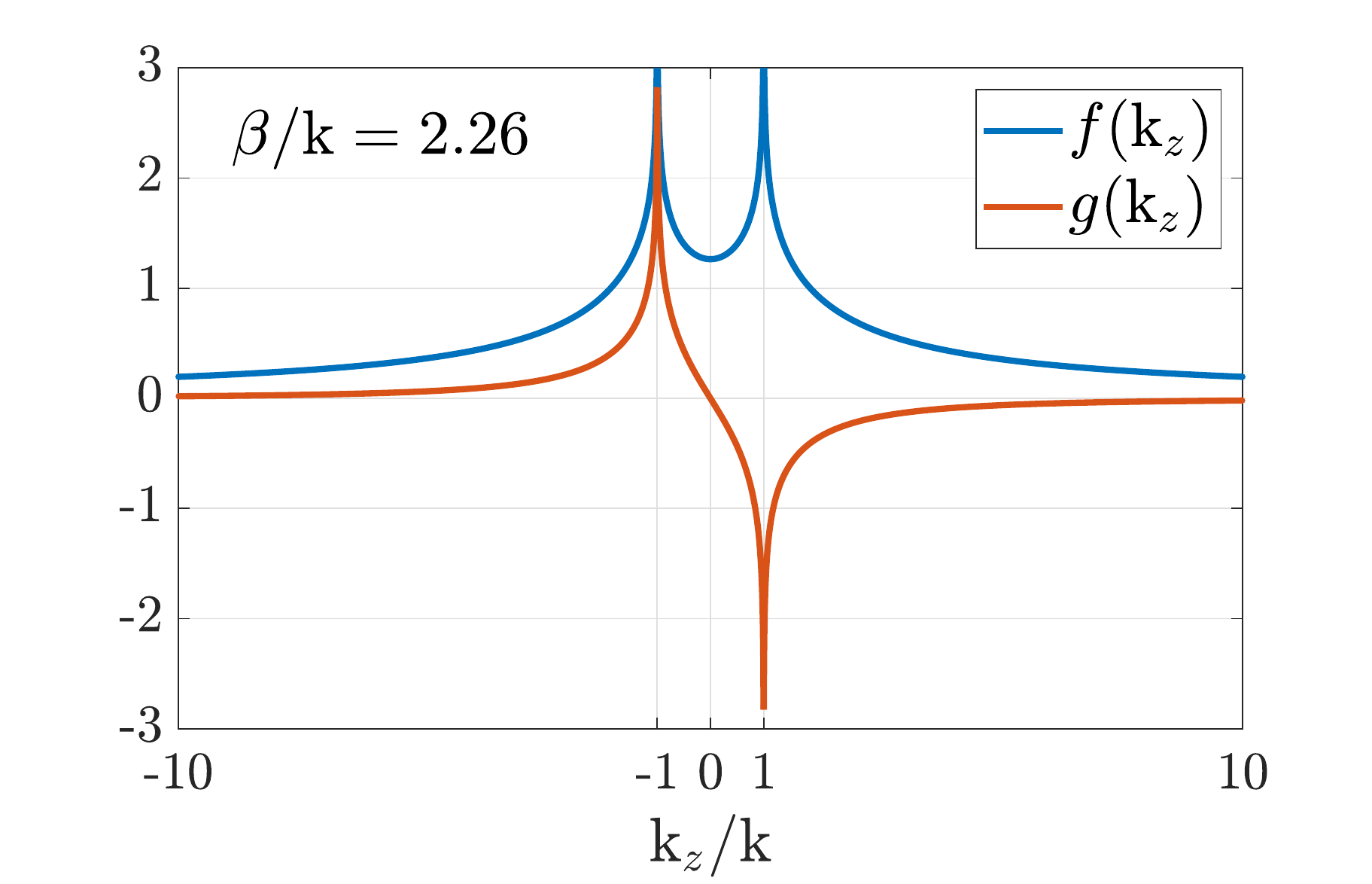}
\caption{\label{fig:fig4} Plot of the functions $f(\mathrm{k}_z),g(\mathrm{k}_z)$ for  $\beta/\mathrm{k}=2.26$.}
\end{figure}

Moreover, it can be shown that the inequality $f(\mathrm{k}_z)\geq |g(\mathrm{k}_z)|\geq0$ always holds true. This has as a consequence the following: \textit{For non-zero $m_z$, the sign of ${R}_{\lambda,m_z j}(\mathrm{k}_z)$ solely depends on the sign of the transverse angular momentum $m_z$, for all $\mathrm{k}_z$}. Additionally, ${R}_{\lambda,m_z j}(\mathrm{k}_z)$ does not depend on the multipolar order $j$, and it has the symmetry property of ${R}_{\lambda,m_z j}(\mathrm{k}_z)={R}_{-\lambda,m_z j}(-\mathrm{k}_z)$.

We will now argue that the $2m_z$-dependent term in Eq~(\ref{eqnPWdirectA}) is the origin of the dominant exponential dependence of the directionality $\restr{D}{(\mathrm{k}, j, m_z, \lambda)}$ on $m_z$. By making use of Eqs.~(\ref{eq:def_coupl}, \ref{eqnTPWS}) and the property of the translation invariance of the system along the x-axis, we can end up with the following representation of the power coupled in the two modes:

\begin{eqnarray}
\restr{\mathrm{C}_{\pm\hat{\mathrm{x}}}}{(\mathrm{k}, j, m_z, \lambda)} &=&\frac{1}{4\pi^2\mathrm{k}^2}\times\label{eqnpowcoupledPW}\\ &\times&\left|\int\limits_{-\infty}^{+\infty}\frac{\mathrm{dk}_{z}}{\mathrm{k}_{y}}\mathrm{A}_{\lambda,m_{z} j}(\hat{\mathrm{k}}_{\pm}) |\hat{\mathrm{e}}_{\lambda}(\hat{\mathrm{k}}_\pm)|\bra{\sigma_{\pm \hat{\mathrm{x}}}} \hspace{2pt} \mathrm{S} \hspace{2pt} \ket{\mathbf{k}_{\pm}\hspace{2pt}\lambda} \right|^2,\nonumber
\end{eqnarray}

\noindent where we represent the normalized plane waves $\hat{\mathrm{e}}_{\lambda}(\hat{\mathrm{k}})/|\hat{\mathrm{e}}_{\lambda}(\hat{\mathrm{k}})|e^{\mathrm{i}\mathbf{k}\cdot\mathbf{r}}$ with the kets $\ket{\mathbf{k}\hspace{2pt}\lambda}$ and also we define $\mathbf{k}_{\pm}(\mathrm{k}_z)=\pm\beta\hat{\mathrm{x}}-\sqrt{\mathrm{k}^2-\beta^2-\mathrm{k}_z^2}\hat{\mathrm{y}}+\mathrm{k}_z\hat{\mathrm{z}}$. One can see from the above equation that the directionality $\restr{D}{(\mathrm{k}, j, m_z, \lambda)}$ will be a function of coherent sums over $\mathrm{k}_z$ of the contributions of all the evanescent components of the multipolar spectrum with $\mathrm{k}_x=\pm \beta$. The cross-section of the waveguide, the multipolar order, and the distance between the emitter and the waveguide will affect the way in which the different $\mathrm{k}_z$-components will be combined. It is not possible to compute $\restr{D}{(\mathrm{k}, j, m_z, \lambda)}$ from our results. For this, one would need to know the S-matrix of the system representing the exact coupling mechanism to the waveguide. However, even though $\restr{D}{(\mathrm{k}, j, m_z, \lambda)}$ is not related directly to ${R}_{\lambda,m_z j}$, using the last line of Eq.~(\ref{eqnPWdirectA}), we can see how the expected trends for it look like. This is because ${R}_{\lambda,m_z j}$, practically, somehow accounts for the elementwise amplitude asymmetry between the two input vectors of the S-matrix of the system that give as outputs the coupling to the two counterpropagating modes. This can be seen by comparing Eqs.~(\ref{eqnratioPW}, \ref{eqnpowcoupledPW}). First, as shown in Sec.~\ref{app:dem}, the overall directionality $\restr{D}{(\mathrm{k}, j, m_z, \lambda)}$ does not depend on helicity $\lambda$ when the system has $\mathrm{M}_z$ mirror symmetry. This means that $\restr{D}{(\mathrm{k}, j, m_z, \lambda)}$ cannot have any $\lambda$-dependent term like the $2\lambda g(\mathrm{k}_z)$ in ${R}_{\lambda,m_z j}$. The other term in ${R}_{\lambda,m_z j}$, with a $2m_z$ dependence, appears for each of the $\mathrm{k}_z$ components, and we therefore expect that $\restr{D}{(\mathrm{k}, j, m_z, \lambda)}$ should show a similar exponential dependence on $m_z$. This expectation is confirmed by the numerical results.

Finally, there is a family of waveguide geometries where $\restr{D}{(\mathrm{k}, j, m_z, \lambda)}$ is directly related with ${R}_{\lambda,m_z j}$. This is the case where, instead of the rectangular waveguide of Fig. 1, we have an arbitrary infinite planar waveguide parallel to the xOz plane. Then, due to the additional translation invariance of such a waveguide along z, the directionality of the coupling of an emitter $\ket{\mathrm{k}\hspace{2pt}j\hspace{2pt}m_z\hspace{2pt}\lambda}$ along its x-axis is given by: $\restr{D}{(\mathrm{k}, j, m_z, \lambda)}={R}_{\lambda,m_z j}(\mathrm{k}_z=0)=2m_z\hspace{1pt}f(\mathrm{k}_z=0)$. Hence, in such a case, the directionality of the coupling of an emitter $\ket{\mathrm{k}\hspace{2pt}j\hspace{2pt}m_z\hspace{2pt}\lambda}$ along the x-axis of the planar waveguide, depends exactly in a proportional way on the transverse angular momentum $m_z$ of the emitter. Moreover, it is independent of helicity $\lambda$, the multipolar order $j$ and the distance between the emitter and the planar waveguide. Apart from its exponential $m_z$-dependence, it only depends on the wavenumber k and the propagation constant $\beta$ of the planar waveguide.
 \end{document}}{}
\end{document}